\documentclass[reprint,prd,amsmath,amssymb,aps,lengthcheck,floatfix,twocolumn,superscriptaddress]{revtex4-1}
\usepackage{balance}
\usepackage{hyperref}
\usepackage{graphicx}
\usepackage{amsfonts}
\usepackage{amssymb}
\usepackage{dcolumn}
\usepackage{bm}
\usepackage{ulem}
\usepackage[utf8]{inputenc}
\usepackage[T1]{fontenc}
\usepackage{amsmath}
\usepackage[version=4]{mhchem}
\usepackage{stmaryrd}
\usepackage{xcolor}
\usepackage{color}
\usepackage{subeqnarray}
\newcommand{\bwt}{\begin{widetext}}
\newcommand{\ewt}{\end{widetext}}
\newcommand{\beq}{\begin{equation}}
\newcommand{\eeq}{\end{equation}}
\newcommand{\bea}{\begin{eqnarray}}
\newcommand{\eea}{\end{eqnarray}}

\begin{document}
\title{Collective modes in relativistic cold asymmetric nuclear matter within the covariant Vlasov approach}
\author{Aziz Rabhi} 
\affiliation{IPEST La Marsa, Carthage University, Tunisia.}
\affiliation{CFisUC, Department of Physics, University of Coimbra, 3004-516 Coimbra, 
Portugal.}
\author{Olfa Boukari} 
\affiliation{ISEP-BG La Soukra, Carthage University, Tunisia.}
\author{Sidney S. Avancini} 
\affiliation{Departamento de F\'{\i}sica, Universidade Federal de Santa Catarina, Florian\'opolis, SC, CP. 476, CEP 88.040-900, Brazil.} 
\author{Constan\c ca Provid\^encia}
\affiliation{CFisUC, Department of Physics, University of Coimbra, 3004-516 Coimbra, Portugal.}

\begin{abstract}
A covariant relativistic approach based on the Vlasov equation is applied to the study of infinite asymmetric nuclear matter. We use several Walecka-type hadronic models and obtain the dispersion relations for the longitudinal modes. The isovector and isoscalar collective modes are determined for a wide range of densities as a function of isospin asymmetry and momentum transfer within a set of eleven relativistic mean field models with different nuclear matter properties. Special attention is given to beta-equilibrium matter. It is shown that the possible propagation of isoscalar and isovector-like modes depends directly on the density dependence of the symmetric nuclear matter equation of state  and of the symmetry energy, with a stiff equation of state  favouring the propagation of isoscalar like collective modes at high densities, and a stiff symmetry energy defining the behavior of the isovector like modes which propagate for densities below two times saturation density. The coupling of the nuclear modes to the electron plasmon is also discussed.
\end{abstract}
\maketitle

\section{Introduction}
A multidisciplinary theoretical effort involving astrophysics, nuclear and particle physics, and statistical physics is required to understand compact stars, supernova cores, and neutron stars. From low densities up to several times nuclear saturation, these stellar objects are essentially composed of neutrons, protons, electrons, muons above a density close to saturation density, and possibly, if their mean free path is short enough, neutrinos. The electrons and muons neutralize the proton charge and thus suppress the divergent Coulomb energy contribution. In addition to understanding the equation of state of the stellar matter, a good description of the mean free path of the neutrinos in the medium is required. Neutrino opacity has been shown to be influenced by nucleon-nucleon interactions through coherent scattering of density fluctuations \cite{Horowitz:2004pv,Horowitz:2004yf}. Both the single particle contribution and the collective contribution must be considered. It is therefore important to understand the collective modes in asymmetric nuclear matter in order to predict the behavior of neutrinos.

Low-density stellar matter consists of neutron-rich nuclei immersed in a gas of neutrons, a consequence of the liquid-gas phase transition that characterizes nuclear matter in this density range.  This matter plays an important role in the collapse of supernovae into neutron stars, see for instance 
\cite{Oertel:2016bki,Hempel:2011mk,Fischer:2013eka}. 
At very low densities, the competition between long-range Coulomb repulsion and short-range nuclear attraction defines the properties of the inhomogeneous matter formed, which is known as nuclear pasta \cite{Ravenhall:1983uh}. It can occur in different structures and its properties have important consequences in the crust of neutron stars and in the core collapse of supernovae \cite{Chamel:2008ca}. It was shown that the isovector channel of nuclear matter  is  important to  define the properties of the transition to the neutron star core \cite{Ducoin:2011fy}. In fact, the curst-core transition may be determined studying the response of nuclear matter to small perturbations, since the appearance of collective unstable modes is intrinsically linked to the existence of a non-homogeneous phase.

In the present work, we are interested in studying the longitudinal nuclear collective stable modes arising from small oscillations around an equilibrium state in nuclear matter using the covariant Vlasov approach \cite{Heinz,avancini2018}. We will  analyze the behavior of several nuclear models used to describe neutron star matter.   The collective modes of nuclear matter have already been studied in Ref.~\cite{Matsui1981} within a Landau-Fermi liquid formalism considering a relativistic mean field description.

In this study both symmetric matter and neutron matter isoscalar modes were discussed. Later, in Ref.~\cite{Lim1989}, a relativistic Hartree calculation was carried out and the zero sound, longitudinal and transverse modes were obtained for symmetric matter. A similar study was undertaken within a semiclassical approach in \cite{nielsen91}, also for symmetric matter. Subsequently, in Refs.~\cite{greco2003,avancini05}, both isoscalar and isovector collective modes were discussed  also within an RMF description.

One of the main goals of this work is to obtain dispersion relations for neutron-proton-electron (NPE) matter within the formalism of the covariant Wigner function, which allows us to calculate the propagation of density modes. The dispersion relations are used to calculate the longitudinal collective modes, both stable and unstable. The unstable modes determine the dynamical spinodal zones that define the boundaries of the inhomogeneous region of the neutron star crust. This formalism is of particular interest because it can be easily generalized to calculate the electrical and thermal conductivity in the magnetized/non-magnetized matter, which is fundamental to the study of neutron star cooling. In a previous work~\cite{avancini2018}, we have studied,  in the context of the covariant Vlasov approach, neutron-proton-electron matter under a strong magnetic field.  In this work, the dispersion relations for the longitudinal and transverse modes were calculated within a  Walecka-type hadronic model and  the instability regions for longitudinal and transverse modes were studied. In particular, the crust-core transition of a magnetized neutron star was discussed in detail.

We investigate the role of isospin and the presence of the Coulomb field and electrons on the longitudinal collective nuclear modes. It has been shown \cite{greco2003,avancini05} that at lower densities an isovector-like collective mode exists. The onset density of the mode depends on the isospin asymmetry. At higher densities, two to three times the saturation density, this mode changes to an isoscalar-like mode and the authors of Ref. \cite{greco2003} have even suggested that the experimental observation of the neutron wave would identify the transition density. We expect the presence of electrons to affect the properties of these modes, namely the excitations with large wavelengths when protons and electrons must move together. We restrict ourselves to the zero-temperature case.

In the following, we will refer to neutral matter composed of protons, neutrons, and electrons as npe matter, and to charged matter composed only of protons and neutrons as np matter. In Sec.~\ref{sectII} we present the covariant Vlasov equation formalism for nuclear matter, including electrons and the electromagnetic field already discussed in Refs. \cite{cp2006, Brito2006}. A brief discussion of the plasmon modes predicted within the present formalism is also included. In Sec. III the numerical results are presented and discussed. Finally, in the last section, the main conclusions are drawn.

%
%
%
%
%
%
%
%
%
\section{Covariant Vlasov equation formalism}
\label{sectII}
For the description of the EoS of neutron star matter, we employ a field-theoretical approach in which the nucleons interact via the exchange of $\sigma$-$\omega$-$\rho$ mesons.  
The Lagrangian density using natural units, i.e., taking $c=\hbar=$1, can be written as
\begin{equation}
{\cal L}=\sum_{j=p,n,e} {\cal L}_j + \cal L_\sigma + {\cal L}_\omega +
{\cal L}_\rho + {\cal L}_{\omega \rho } + {\cal L}_{A} \ ,
\label{lag}
\end{equation}
with
\begin{equation}
{\cal L}_j=\bar \psi^{(j)}\left[\gamma^\mu i D_\mu^{(j)}-M^*_j \right]\psi^{(j)} ,\nonumber
\end{equation}
where the covariant derivative 
is defined as, $ iD^{(j)}_{\mu}~= ~ i \partial_{\mu} - 
{\cal V}^{(j)}_{\mu} $,
where $j=(n,p,e)$, stands for the neutron, proton and electron,
\begin{equation}
 {\cal V}^{(j)}_{\mu} = 
 \left\{
       \begin{array}{l}
         g_v V_\mu  + \frac{g_\rho}{2}\, \vec{b}_\mu+ e\, A_\mu \ ~ ,~j=p \\  
         g_v {V}_\mu -\frac{g_\rho}{2}\, \vec{b}_\mu \ ~ ,~j=n \\
         - e\, A_\mu \ ~ ,~ j=e
     \end{array}    \right. \ ,  \label{eq2}
\end{equation}
$M^*_p=M^*_n=M^*=M-g_s\phi(x),$ $M^*_e=m_e$ and
$e=\sqrt{4\pi/137}$ is  the electromagnetic coupling constant. In order to study the effect of the density dependence of the nuclear symmetry energy on the collective modes in relativistic cold asymmetric nuclear matter, we choose two sets of models  satisfying constraints from Multimessenger Resonant Shattering Flares~\cite{Duncan2023}.
For the nuclear matter parameters, we will consider the following set of RMF models: 
FSU \cite{fsu} and FSU2 \cite{fsu2}, FSU2H and FSU2R \cite{tolos17,tolos17_2,Negreiros18}, NL3
\cite{nl3} and NL3 $\omega\rho$ \cite{Pais16, Horowitz01}, TM1 \cite{tm1},  TM1e \cite{tm1e}, TM1-2 and TM1-2 $\omega\rho$ \cite{tm1-2}, and BigApple \cite{bigapple} (the full set of parameters of the models can be found in Table \ref{tab:parameters}, and for their properties see Table \ref{tab:nuclear}).
\begin{table*}[htb]
 \begin{tabular}{c|ccccc|cccccc}
\hline
\hline
&&& Set I&&&&& Set II\\
  & NL3 & NL3$\omega\rho$ & TM1-2 & TM1-2$\omega\rho$ & BigApple & FSU & FSU2 & FSU2R & FSU2H & TM1 & TM1e \\
\hline 
$\rho_0$ $ [\mathrm{fm}^{-3}]$ & 0.148 & 0.148 & 0.145 & 0.146 & 0.155 & 0.148 & 0.1505 & 0.1505 & 0.1505 & 0.145 & 0.145\\
$m_\sigma$ (MeV) & 508.194 & 508.194 & 511.198 & 511.198 & 492.730 & 491.500 & 497.479 & 497.479 & 497.479 & 511.198  & 511.198\\
$m_\omega$ (MeV) & 782.501 & 782.501 & 783.000 & 783.000 & 782.500 & 782.500 & 782.500 & 782.500 & 782.500 & 783.000  &783.000 \\
$m_\rho$ (MeV)   & 763.000 & 763.000 & 770.000 & 770.000& 763.000 & 763.000 & 763.000 & 763.000 & 763.000 & 770.000  & 770.000\\
$g_{\sigma}$     &  10.217 &  10.217 & 9.998  & 9.998  &   9.670 &  10.592 &  10.397 &  10.372 &  10.135 &  10.0289 & 10.029\\
$g_{\omega}$     &  12.868 &  12.868 & 12.503 & 12.503 &  12.316 &  14.302 &  13.557 &  13.505 &  13.020 &  12.6139 & 12.614\\
$g_{\rho}$       &   8.948 &  11.277 & 9.264  & 11.302 &  14.162 &  11.767 &   8.970 &  14.367 &  14.045 &   9.2644 & 13.971\\
$\kappa$         &  4.3840 &  4.3840 & 3.5235 &  3.5235 & 5.0105  & 1.7976 & 3.5940  &  3.6729 & 4.4364 & 3.0397 &  3.0397  \\
$\lambda$        & -173.31 & -173.31 & -47.36 & -47.36  & -190.08 & 299.13  & -6.23  & -19.44  & -140.31 & 3.7098 & 3.7098  \\
$\xi$            & 0.0000 & 0.0000 & 0.011267 & 0.011267 & 0.00070 & 0.0600 & 0.0256 & 0.024 & 0.008 & 0.0169 & 0.0169 \\
$\Lambda_{v}$ & 0.0000 & 0.0300 & 0.0000 & 0.0300 & 0.047471 & 0.0300 & 0.000823 & 0.0450  &  0.0450 & 0.0000 & 0.0429 \\
\hline
\end{tabular}
\caption{Coupling parameters of the RMF models considered in this study. \label{tab:parameters}
}
\end{table*}
The meson and photon contributions in Eq.~(\ref{lag}) are given by
\begin{eqnarray}
\mathcal{L}_{{\sigma }} &=&\frac{1}{2}\left( \partial _{\mu }\phi \partial %
^{\mu }\phi -m_{s}^{2}\phi ^{2}-\frac{1}{3}\kappa \phi ^{3}-\frac{1}{12}%
\lambda \phi ^{4}\right) \ , \nonumber \\
\mathcal{L}_{{\omega }} &=&\frac{1}{2} \left(-\frac{1}{2} \Omega _{\mu \nu }
\Omega ^{\mu \nu }+ m_{v}^{2}V_{\mu }V^{\mu }
+\frac{1}{12}\xi g_{v}^{4}(V_{\mu}V^{\mu })^{2} 
\right) \ , \nonumber \\
\mathcal{L}_{{\rho }} &=&\frac{1}{2} \left(-\frac{1}{2}
{\vec{B}}_{\mu \nu }\cdot {\vec{B}}^{\mu
\nu }+ m_{\rho }^{2} \vec{b}_{\mu }\cdot \vec{b}^{\mu } \right)   \ , \nonumber \\
\mathcal{L}_{\omega \rho } &=& \Lambda_v g_v^2 g_\rho^2 V_{\mu }V^{\mu }
\vec{b}_{\nu }\cdot \vec{b}^{\nu }   \nonumber \\
\mathcal{L}_{A} &=&-\frac{1}{4} F_{\mu \nu }F ^{\mu \nu }~, \label{mesonlag}
\end{eqnarray}
where $\Omega _{\mu \nu }=\partial _{\mu }V_{\nu }-\partial _{\nu }V_{\mu }$, 
$\vec{B}_{\mu \nu }=\partial _{\mu }\vec{b}_{\nu }-\partial _{\nu }
\vec{b}
_{\mu }-\Gamma_{\rho }(\vec{b}_{\mu }\times \vec{b}_{\nu })$ and 
$F_{\mu \nu }=\partial _{\mu }A_{\nu }-\partial _{\nu }A_{\mu }$.
The parameters $\kappa$, $\lambda$ and $\xi$ are self-interacting couplings and the $\omega$-$\rho$ coupling $\Lambda_v$ is included in order to soften the density dependence of the symmetry energy above the saturation density.
From the Euler-Lagrange equations, one can obtain the Dirac equation for the fields of the fermions:
\begin{eqnarray}
 i \gamma^\mu D^{(j)}_{\mu}~\psi^{(j)} = M^{\star}_{j} ~\psi^{(j)} \ , \label{dirac1}
\end{eqnarray}
and its conjugate equation:
\begin{eqnarray}
 \bar{\psi}^{(j)} i D^{\dagger (j)}_{\mu}\gamma^\mu ~= - M^{\star}_{j} ~\bar{\psi}^{(j)} \, 
 \label{dirac2}
\end{eqnarray}
where $ iD^{\dagger (j)}_{\mu} = ~ i\overleftarrow{\partial}_{\mu} + {\cal V}^{(j)}_{\mu} $ .
In this section we discuss how the Vlasov equation for a hadronic system is obtained from general transport equations. Our formalism is based on the covariant Wigner function described in detail in Ref. \cite{avancini2018}. In this section, we will present only the main results related to transport theory in order to keep this paper minimally self-contained. We will focus on the new technical details that arise when using the covariant Wigner function to describe npe matter. The present formalism is suitable for the study of collective modes and opens the possibility to study the thermal and electrical conductivity in npe matter.

\subsection{Covariant Vlasov approach}
We start from the generalized Vlasov equation established in \cite{avancini2018}, given by:
 \begin{equation}
\partial_t f_{(j)} + \vec{v} \cdot \nabla_x f_{(j)} +
  (\vec{{\cal E}} + \vec{v} \times \vec{\cal B})\cdot \nabla_p f_{(j)} = 0, \quad j=n, p, e,
  \label{vlasov1}
 \end{equation}
where $\vec{v}=\vec{p}/E_p$ and
\begin{eqnarray}
 && {\cal E}_i^{(j)} = {\cal F}_{0i}^{(j)} - \frac{M_j^\star(x)}{E_p^{(j)}} 
 \nabla_{x,i}~ M_j^\star(x) ~\ , \  j=p,n \nonumber \ , \\
 && {\cal E}_i^{(j)} = - \partial_t ~\delta \vec{{\cal V}}^{(j)} -
 \nabla_x \delta {\cal V}_0^{(j)} - \frac{M_j^\star(x)}{E_p^{(j)}} 
 \nabla_{x,i}~ M_j^\star(x),~\, j=p,n \nonumber \, \\
 && {\cal E}_i^{(e)} = {\cal F}_{0i}^{(e)}  \nonumber \ , \\
 && {\cal B}^{(j)}_i ~=~\epsilon_{ilm} \partial_{x,l} {\cal V}^{(j)}_m ~,~j=p,n,e~\,
\end{eqnarray}
$i,l,m=1,2,3$, and $f^{(j)}$  is the distribution function of j-specie.

In order to obtain the dispersion relations we analyze the current densities of baryons and electrons,
$$
\begin{aligned}
 J_{\mu}(x)&=\sum_{j=n,p,e}\frac{2}{(2\pi)^3}\int \frac{d^3 p}{p^0} p_{\mu} f^{(j)}(x, \vec{p}) \\
 & = \sum_{j=n,p,e} J_{\mu}^{(j)} (x), \quad p^{0}=E_{j}^{(0)}=\sqrt{\vec{p}^2+{M^{* (0)}_j}^2}.
\end{aligned}
$$
where $M_j^{\star (0)}= M-g_s \phi^{(0)}$ and $M_e^{\star (0)}$=$m_e$. 
From the generalized Vlasov equation,~Eq.(\ref{vlasov1}), we obtain the following conservation law: $\partial^{\mu}J^{(j)}_{\mu}=0$.

We will consider small oscillations in relation to the equilibrium generated by perturbations of the distribution function
\begin{equation}
 f^{(j)}(\vec{p}) =f^{(0)(j)}(\vec{p})+\delta f^{(j)}(\vec{p}).
 \label{delphi}
\end{equation}
Since in the present work we are only interested in systems at zero temperature, the equilibrium distribution function is given by 
\beq
f^{(0)(j)}(\vec{p})=\theta\left(p^2_{Fj} -p^2 \right), j=p, n, e,
\label{wignereq5}
\eeq 
where the Heaviside function $\theta(x)$ was used.

The small perturbation of the distribution functions, $f^{(0)(j)}$, around their equilibrium 
values, will generate perturbations on the fields:
\begin{eqnarray}
&& \phi = \phi^{(0)}+\delta \phi ~,~V_\mu=V_\mu^{(0)}+\delta V_\mu ~,~
 b_\mu=b_\mu^{(0)}+\delta b_\mu~, \nonumber \\
 &&~A_\mu=A_\mu^{(0)}+ \delta A_{\mu}~ , \label{smalldev}
\end{eqnarray}
and cause a corresponding perturbation of the equilibrium 4-current,
\begin{equation}
  J^{(j)}_\mu(x) = J^{(0)(j)}_\mu
   + \delta J^{(j)}_\mu \ ,
\end{equation}
with
\begin{equation}
  \delta J^{(j)}_\mu = \frac{2}{(2\pi)^3} \int \frac{d^3 p}{E^{(0)}_j}
    ~p_\mu ~\delta f^{(j)}. \label{curre}
\end{equation}
The scalar density is given by the expression:
 \begin{equation}
  \rho_s^{(j)}= \frac{2}{(2\pi)^3} \int \frac{d^3 p}{E_j} {M_j^\star}
 f^{(j)}(x,\vec{p})    \ .
 \end{equation}
The small perturbation of the proton and neutron scalar densities have to be calculated with 
care \cite{avancini05}, since ${M_j^\star}= M-g_s \phi(x)$ is position-dependent, 
resulting in the following expression:
\begin{equation}
 \rho_s^{(j)}= \rho_s^{(0)(j)}+ \delta \rho_s \ ,
\end{equation}
with
\begin{equation}
  \rho_s^{(0)(j)}= \frac{2}{(2\pi)^3} \int \frac{d^3 p}{E^{(0)}_j} {M_j^\star}^{(0)} 
 f^{(0)(j)}(\vec{p})    \ , 
\end{equation}
and $ \delta \rho_s^{(j)}= \delta\tilde{\rho}_s^{(j)}+ g_s ~d\rho_s^{(0)(j)}~ \delta \phi$ ,
with:
 \begin{equation}
  \delta \tilde{\rho}_s^{(j)}= \frac{2}{(2\pi)^3} \int \frac{d^3 p}{E^{(0)}_j} 
  {M_j^\star}^{(0)} \delta f^{(j)}    \ ,  \label{densca}
 \end{equation}
 and
 \begin{equation}
  d\rho_s^{(0)(j)} = - \frac{2}{(2\pi)^3} \int d^3 p 
  \frac{{\vec{p}}^2} {{E^{(0)}_j}^3}   f^{(0)(j)}    ~.
 \end{equation}
\noindent
After substituting Eq.(\ref{delphi}) in the Vlasov equation, Eq.(\ref{vlasov1})
retaining only terms of the first order in $\delta f^{(j)}$, one obtains:
 \begin{eqnarray}
  && \partial_t \delta f^{(j)} + \vec{v} \cdot \nabla_x \delta f^{(j)} +
  \vec{v} \times ( \nabla_x \times \vec{\cal V}^{(0)(j)} ) \cdot \nabla_p \delta f^{(j)} 
  \cr
  &&
  + \left[ \vec{v} \times \nabla_x \times (\vec{\cal V}^{(0)(j)}  + 
\delta \vec{\cal V}^{(j)} ) + g_s \frac{{M_j^\star}^{(0)}}{{E^{(0)}_j}} \nabla_x \delta \phi 
\right.  \nonumber \\ &&  \left.   
- \partial_t ~\delta \vec{{\cal V}}^{(j)} -
 \nabla_x \delta {\cal V}_0^{(j)}  \right]  \cdot 
 \nabla_p f^{(0)(j)}    
   = 0 \ ,
  \label{dvlasov}
 \end{eqnarray}
where $\vec{v}=\vec{p}/E^{(0)}_j$ , $j=p,e$ ( for the electrons $g_s=0$). 
The last equation can be further simplified noting that the equilibrium distribution 
function, eq. (\ref{wignereq5}), hence,
it is useful to rewrite eq. (\ref{dvlasov}) as:
 \bea
&&\partial_t \delta f^{(j)} + \vec{v} \cdot \nabla_x \delta f^{(j)}
+ \left[ \vec{v}  \times (\nabla_x \times \delta \vec{\cal V}^{(j)} ) + 
  g_s \frac{{M_j^\star}^{(0)}}{{E^{(0)}_j}} \nabla_x \delta \phi \right.\cr
&&
\left. - \partial_t ~\delta \vec{{\cal V}}^{(j)} -\nabla_x \delta {\cal V}_0^{(j)}  
 \right]  \cdot 
 \nabla_p f^{(0)(j)}    
   = 0 \ .
  \label{dvlasov2}
 \eea
Next, we obtain the dispersion relations, starting from the Fourier transform
of the small deviation from equilibrium of the fields and of the distribution functions:
 \begin{equation}
 \left\{
       \begin{array}{c}
          \delta f^{(j)} (\vec{x},\vec{p},t) \\
         \delta \phi (\vec{x},t) \\
         \delta {\cal V}^{(j)}_{\mu} (\vec{x},t)
       \end{array}    
 \right\}  = 
 \int d^3q ~d\omega 
  \left\{
       \begin{array}{c}
          \delta f^{(j)} (\vec{q},\omega,\vec{p}) \\
          \delta \phi (\vec{q},\omega) \\
          \delta {\cal V}^{(j)}_{\mu}(\vec{q},\omega) 
       \end{array}    
 \right\}  e^{i(\omega t - \vec{q} \cdot \vec{x})} \ ,
 \end{equation}
and after substituting the last equation in the Vlasov equation, eq. (\ref{dvlasov2}), one 
obtains for $ \delta f^{(j)}(\vec{q},\omega,\vec{p})$ :
\bwt
 \begin{equation}
 i (\omega - \vec{v} \cdot \vec{q} ) ~\delta f^{(j)} 
 = i \left[ (\omega - \vec{v} \cdot \vec{q} ) ~\delta \vec{\cal V}^{(j)}  -
  \left( \delta {\cal V}_0^{(j)} - \vec{v} \cdot \delta \vec{{\cal V}}^{(j)}
    - g_s \frac{{M_j^\star}^{(0)}}{{E^{(0)}_j}} \delta \phi \right) \vec{q}  
 \right]  \cdot 
 \nabla_p f^{(0)(j)}.
  \label{dvlasov3}
 \end{equation}
\ewt

It is important to highlight that the Vlasov equation can be derived through various methods, for instance, in references \cite{nielsen89} and \cite{nielsen91} Nielsen et al. utilize a Hamiltonian formalism to obtain this equation. The Wigner approach offers a significant advantage by providing a systematic method for calculating particle equilibrium distribution functions, which are essential for accurately describing systems in a magnetized medium.
For the small deviation from equilibrium of the distribution function, the generalized Vlasov equation is given by, as detailed in Eq.~(37) from our previous paper Ref.\cite{avancini2018}:
\bwt
\begin{equation}
\delta f^{(j)} = \left[\delta \overrightarrow{\mathcal{V}}^{(j)}-\frac{\left(\delta\mathcal{V}^{(j)}_0-\frac{\vec{p} \cdot \delta \overrightarrow{\mathcal{V}}^{(j)}}{E_{j}^{(0)}}-g_{s} \frac{M_{j}^{\star(0)}}{E_{j}^{(0)}} \delta \phi\right)}{\omega-\frac{\vec{p} \cdot \vec{q}}{E_{j}^{(0)}}} \vec{q}\right] \cdot \nabla_{p} f^{(0)}_{(j)}.
\label{vlasov}
\end{equation}
\ewt
where $\delta f^{(j)}, \mathcal{V}_{\mu}^{(j)}$, and $\delta \phi$ are functions of $(\vec{q}, \omega, \vec{p})$.

For convenience, we adopt the reference frame where $\vec{q}=q_{z} \vec{k}$ and, we define the following expressions:
$$
\begin{aligned}
& \vec{q}=q_{z} \vec{k} \equiv \vec{q}_{\|}=q \hat{k} \quad, \quad \vec{p}=p_{x} \hat{i}+p_{y} \hat{j}+p_{z} \hat{k}=\vec{p}_{\perp}+\vec{p}_{\|}, \\
& \vec{p}_{\perp}=p_{x} \hat{i}+p_{y} \hat{j} \quad, \quad \vec{p}_{\|}=p_{z} \hat{k}, \\
& \vec{p}=p \sin \theta \cos \Phi \hat{i}+p \sin \theta \sin \Phi \hat{j}+p \cos \theta \hat{k}, \\
& \vec{p}=p \hat{p} \quad, \quad \hat{p}=\sin \theta \cos \Phi \hat{i}+\sin \theta \sin \Phi \hat{j}+\cos \theta \hat{k},\\
& \vec{p}=p \hat{p} \quad, \quad \hat{p}=\sin \theta\hat{e}_{\perp} +\cos \theta \hat{e}_{\|}.
\end{aligned}
$$
\subsection{DISPERSION RELATIONS FOR LONGITUDINAL MODE}

We consider the particular case of small perturbations that correspond to longitudinal waves in our reference frame. The longitudinal mode corresponds to small perturbations parallel to $\hat{k}$ and the dispersion relations are obtained by taking $q_{\perp}=0, \; q_{\|} =q$ and $\delta \mathcal{V}_{x}=\delta \mathcal{V}_{y}=0$.
$$
\delta \overrightarrow{\mathcal{V}}^{(j)}=\delta \overrightarrow{\mathcal{V}}_{\|}^{(j)} \equiv \delta \mathcal{V}_{z}^{(j)} \hat{k}, \quad \delta \mathcal{V}_{x}^{(j)}=\delta \mathcal{V}_{y}^{(j)}= \delta \mathcal{V}_{\perp}^{(j)}=0
$$
and,
$$
\delta \overrightarrow{J}^{(j)}=\delta \overrightarrow{J}_{\|}^{(j)} \equiv \delta J_{z}^{(j)} \hat{k}, \quad \delta J_{\perp}^{(j)}=0
$$
From the conservation law, $\partial^{\mu} J_{\mu}^{(j)}=0$, using the Fourier transformation
$\partial^{\mu} J_{\mu}^{(j)}=\partial^{\mu} J_{\mu}^{(j)}(t,\; \vec{x})=0$ for $j=n, p, e$ leads to the following relation
$$\omega J_0^{(j)}(\vec{q}, \omega)=\vec{q}\cdot\vec{J}^{\;(j)}(\vec{q}, \omega)$$
then,
$$\omega \delta J^{(j)}_0(\vec{q}, \omega)=\vec{q}\cdot\delta \vec{J}^{(j)}(\vec{q}, \omega)=q \delta J_{z}^{(j)}(\vec{q}, \omega).$$
The equations of motion for the mesons and the electromagnetic fields follow
from the use of the Euler-Lagrangian equations in the meson Lagrangian:
\begin{eqnarray}
 && \partial_t^2 \phi -\nabla^2 \phi + m_s^2 \phi +  \frac{\kappa}{2} \phi^2 +
 \frac{\lambda}{6} \phi^3 = g_s \sum_{j=p,n} \rho_s^{(j)}  \nonumber \\
 && \partial_t^2 V_{\mu} -\nabla^2 V_{\mu} + m_v^2 V_{\mu} + \frac{\xi}{6} V_\nu V^\nu V_\mu +
 2 \Lambda_v b_\nu b^\nu V_\mu
  = g_v \sum_{j=p,n} J^{(j)}_\mu  \nonumber \\
 &&  \partial_t^2 b_{\mu} -\nabla^2 b_{\mu} + m_\rho^2 b_{\mu} +
 2 \Lambda_v V_\nu V^\nu b_\mu
  = \frac{g_\rho}{2} \sum_{j=p,n} \tau_j J^{(j)}_\mu  \nonumber \\ 
  &&  \partial_t^2 A_{\mu} -\nabla^2 A_{\mu} 
  = e( J^{(p)}_\mu - J^{(e)}_\mu) =  \sum_{j=p,e} Q_j J^{(j)}_\mu   \ . 
\end{eqnarray} 
Now, we consider small deviations from the equilibrium in the fields as given in Eq.(\ref{smalldev}) and perform a Fourier transform obtaining:
\begin{eqnarray}
&& \left[ -\omega^2 + {\vec q}^{~2} + {\tilde m}_s^2 \right] \delta \phi (\vec{q},\omega)
= g_s \sum_{j=p,n} \frac{2 {M^\star}^{(0)(j)}}{(2\pi)^3}\int \frac{d^3 p}{E^{(0)}_j}\delta f^{(j)},  \cr
&&\left[ -\omega^2 + {\vec q}^{~2} + m_v^2 + \frac{\xi}{6}{V^{(0)}_0}^2 +
 2 \Lambda_v {b^{(0)}_0}^2 \right] \delta V_{\mu} +\frac{\xi}{3} {V_0^{(0)}}^2 \delta V_0 \delta_{\mu 0} \cr
&&+ 4 \Lambda_v V_0^{(0)} b^{(0)}_0 \delta b_0 \delta_{\mu 0}
  = g_v  \sum_{j=p,n} \frac{2}{(2\pi)^3}\int \frac{d^3 p}{E^{(0)}_j} p^\mu \delta f^{(j)},\cr
&& \left[ -\omega^2 + {\vec q}^{~2} + m_\rho^2 +
             2 \Lambda_v {V^{(0)}_0}^2 \right] \delta b_{\mu} \cr
&& + 4 \Lambda_v V_0^{(0)} b^{(0)}_0 \delta V_0\delta_{\mu 0}
  = \frac{g_\rho}{2}  \sum_{j=p,n} \tau_j \frac{2}{(2\pi)^3}\int \frac{d^3 p}{E^{(0)}_j} p^\mu \delta f^{(j)},  \cr
&& \left[ -\omega^2 + {\vec q}^{~2} \right] \delta A_{\mu} = \sum_{j=p,e} Q_j \frac{2}{(2\pi)^3}
  \int \frac{d^3 p}{E^{(0)}_j} p^\mu \delta f^{(j)}, 
  \label{eomf}
\end{eqnarray}
where the effective scalar mass is given by:
\begin{equation}
 {\tilde m}_s^2 = m_s^2 + \kappa \phi^{(0)} + \frac{\lambda}{2}{\phi^{(0)}}^2 - g_s^2
 \sum_{j=p,n} d \rho^{(0)(j)}_s \ .
\end{equation}
 \begin{equation}
  d\rho_s^{(0)(j)} = - \frac{2}{(2\pi)^3} \int d^3 p 
  \frac{{\vec{p}}^2} {{E^{(0)}_j}^3}   f^{(0)(j)}~.
 \end{equation}
Next, we use the definition of the current,
\beq
\delta J_{\mu}^{(j)}(\vec{q}, \omega)=\frac{2}{(2 \pi)^{3}} \int \frac{d^{3} p}{E_{j}^{(0)}} p^{\mu} \delta f^{(j)}(\vec{q}, \omega, \vec{p}), \quad j \in(p, n, e)
\eeq
and write the non-null components of the meson vector field fluctuations as,
\bwt
\bea
&& {\left[-\omega^{2}+\vec{q}^{\;2}+m_{v}^{2}+\frac{\xi}{2} V_{0}^{(0)^{2}}+2 \Lambda_{v} b_{0}^{(0)^{2}}\right] \delta V_{0} +4 \Lambda_{v} V_{0}^{(0)} b_{0}^{(0)} \delta b_{0}=g_{v} \sum_{j=p, n} \delta J_{0}^{(j)}} \cr
&& {\left[-\omega^{2}+\vec{q}^{\;2}+m_{v}^{2}+\frac{\xi}{6} V_{0}^{(0)^{2}}+2 \Lambda_{v} b_{0}^{(0)^{2}}\right] \delta V_{z}=g_{v} \sum_{j=p, n} \delta J_{z}^{(j)}} \cr
&& {\left[-\omega^{2}+\vec{q}^{\;2}+m_{\rho}^{2}+2 \Lambda_{v} V_{0}^{(0)^{2}}\right] \delta b_{0}+4 \Lambda_{v} V_{0}^{(0)} b_{0}^{(0)} \delta V_{0}=\frac{g_{\rho}}{2} \sum_{j=p, n} \tau_{j} \delta J_{0}^{(j)}} \cr
&& {\left[-\omega^{2}+\vec{q}^{\;2}+m_{\rho}^{2}+2 \Lambda_{v} V_{0}^{(0)^{2}}\right] \delta b_{z}=\frac{g_{\rho}}{2} \sum_{j=p, n} \tau_{j} \delta J_{z}^{(j)}},
\eea
\ewt
where, the effective mesons masses are given by,
\bea
\tilde{m}_{\omega}^{2}&=&m_{v}^{2}+\frac{\xi}{2} V_{0}^{(0)^{2}}+2 \Lambda_{v} b_{0}^{(0)^{2}} \cr
\tilde{m}_{\rho}^{2}&=&m_{\rho}^{2}+2 \Lambda_{v} b_{0}^{(0)^{2}}. 
\eea
We use a new definition,
\begin{eqnarray}
\omega^2_s &=&  {\tilde m}_s^2 + {\vec q}^{~2} \cr
\omega^2_{\omega} &=& {\tilde m}_{\omega}^2 + {\vec q}^{~2} \cr
\omega^2_{\rho} &=& {\tilde m}_{\rho}^2 + {\vec q}^{~2}
\end{eqnarray}
The components of the vector field fluctuations can be written as
\bea
&& {\left[-\omega^{2}+\omega_{\omega}^{2}\right] \delta V_{0}+4 \Lambda_{v} V_{0}^{(0)} b_{0}^{(0)} \delta b_{0}=g_{v} \sum_{j=p, n} \delta J_{0}^{(j)}} \cr
&& {\left[-\omega^{2}+\omega_{\omega}^{2}-\frac{\xi}{3} V_{0}^{(0)^{2}}\right] \delta V_{z}=g_{v} \sum_{j=p, n} \delta J_{z}^{(j)}} \cr
&& {\left[-\omega^{2}+\omega_{\rho}^{2}\right] \delta b_{0}+4 \Lambda_{v} V_{0}^{(0)} b_{0}^{(0)} \delta V_{0}=\frac{g_{\rho}}{2} \sum_{j=p, n} \tau_{j} \delta J_{0}^{(j)}} \cr
&& {\left[-\omega^{2}+\omega_{\rho}^{2}\right] \delta b_{z}=\frac{g_{\rho}}{2} \sum_{j=p, n} \tau_{j} \delta J_{z}^{(j)}}
\label{EQ6}
\eea
The temporal component of the fields,
\bea
\left[-\omega^{2}+\omega_{\omega}^{2}\right] \delta V_{0}+4 \Lambda_{v} V_{0}^{(0)} b_{0}^{(0)} \delta b_{0}&=&g_{v} \sum_{i=p, n} \delta \rho_{i} \cr
\left[-\omega^{2}+\omega_{\rho}^{2}\right] \delta b_{0}+4 \Lambda_{v} V_{0}^{(0)} b_{0}^{(0)} \delta V_{0}&=&\frac{g_{\rho}}{2} \sum_{i=p, n} \tau_{i} \delta \rho_{i}\quad 
\eea
We solve the latter equations for $\delta V_{0}$ and $\delta b_{0}$,
\bwt
\bea
\delta V_{0}&=&\frac{\left[g_{v}\left(-\omega^{2}+\omega_{\rho}^{2}\right) \sum_{i=p, n} \delta \rho_{i}-4 \Lambda_{v} V_{0}^{(0)} b_{0}^{(0)} \frac{g_{\rho}}{2} \sum_{i=p, n} \tau_{i} \delta \rho_{i}\right]}{\left(-\omega^{2}+\omega_{\omega}^{2}\right)\left(-\omega^{2}+\omega_{\rho}^{2}\right)-\left(4 \Lambda_{v} V_{0}^{(0)} b_{0}^{(0)}\right)^{2}}  \\
 \delta b_{0}&=&\frac{\left[\frac{g_{\rho}}{2}\left(-\omega^{2}+\omega_{\omega}^{2}\right) \sum_{i=p, n} \tau_{i} \delta \rho_{i}-4 \Lambda_{v} V_{0}^{(0)} b_{0}^{(0)} g_{v} \sum_{i=p, n} \delta \rho_{i}\right]}{\left(-\omega^{2}+\omega_{\omega}^{2}\right)\left(-\omega^{2}+\omega_{\rho}^{2}\right)-\left(4 \Lambda_{v} V_{0}^{(0)} b_{0}^{(0)}\right)^{2}} 
 \label{fieldt}
\eea
\ewt
The conservation of the current gives:
$
\partial^{\mu} \delta J_{\mu}^{(j)}=0 \rightarrow \omega \delta J_{0}^{(j)}-q \delta J_{z}^{(j)}=0 \quad \rightarrow \quad \delta J_{z}^{(j)}=\frac{\omega}{q} \delta J_{0}^{(j)}
$
and using the conservation of current in Eq.~\ref{EQ6} to obtain the spatial component of the fields:
\bea
\delta V_{z}&=&\frac{(\frac{\omega}{q})}{-\omega^{2}+\omega_{\omega}^{2}-\frac{\xi}{3} V_{0}^{(0)^{2}}} \; g_{v} \sum_{i=p, n} \delta \rho_{i} \cr
\delta b_{z}&=&\frac{(\frac{\omega}{q}) }{-\omega^{2}+\omega_{\rho}^{2}}\frac{g_{\rho}}{2} \sum_{i=p, n} \tau_{i} \delta \rho_{i}
\label{fields}
\eea
Next, we rewrite the Vlasov equation, Eq.(\ref{vlasov}), for the description of longitudinal mode, and rearrange it as follows:
\bwt
\beq
\delta f^{(j)} =\delta\mathcal{V}^{(j)}_{z}\left(\frac{\partial f^{(0)(j)}}{\partial p_{z}}+\frac{p_{z}}{E^{(0)}_{j}}\frac{\vec{q}\cdot \nabla_{p} f^{(0)(j)}}{\left(\omega-\frac{\vec{p}_{z} \cdot \vec{q}}{E_{j}^{(0)}}\right)} \right)
-\frac{\delta\mathcal{V}^{(j)}_0\vec{q} \cdot \nabla_{p} f^{(0)(j)}}{\left(\omega-\frac{\vec{p}_{z} \cdot \vec{q}}{E_{j}^{(0)}}\right)}
+\frac{M_{j}^{\star(0)} g_{s} \delta \phi \; \vec{q} \cdot \nabla_{p} f^{(0)(j)}}{E_{j}^{(0)}\left(\omega-\frac{\vec{p}_{z} \cdot \vec{q}}{E_{j}^{(0)}}\right)}, \quad j=p, n.
\label{vlasovL}
\eeq
\ewt
We highlight an important point here: when the self-interaction of $\omega$ and $\omega-\rho$ interaction terms are incorporated into the model Lagrangian, current conservation no longer yields a simple relationship between the temporal and spatial components of the vector fields:
$$
\delta V_{z}=\frac{\omega}{q} \delta V_{0} \quad, \quad \delta b_{z}=\frac{\omega}{q} \delta b_{0}
$$
The latter relation, which is usually used and simplifies the calculation of the dispersion relations, is only valid when self-interaction $\omega$ and $\omega-\rho$ interactions are absent. It is easy to see that the latter simple relations are recovered when $\xi=\Lambda_{v}=0$ in Eqs.~(\ref{fieldt}, \ref{fields}). It is trivial to show that the current conservation for the electron current yields the usual relation:
$$
\delta A_{z}=\frac{\omega}{q} \delta A_{0}
$$

From the Vlasov formalism, Eq.~\ref{vlasovL}, the dispersion relation for density perturbations can be obtained as following,
\bea
 \delta\rho_j\equiv\delta J_{0}^{(j)}&=& \frac{2}{(2\pi)^3}\int d^3 p\, \delta f^{(j)}\cr
 & =&\frac{1}{2\pi^2}p_{F j}E^{(0)}_{Fj}L(s_j)\left(\frac{\omega}{q}\delta\mathcal{V}^{(j)}_{z}-\delta\mathcal{V}^{(j)}_{0}\right) \cr
 && +\frac{1}{2\pi^2}p_{F j}L(s_j)M_{j}^{\star(0)} g_{s} \delta \phi
 \label{eq37}
\eea
where, 
$s_{j}=\frac{\omega}{\omega_{0j}},\quad\omega_{0j}=\frac{qp_{F_j}}{E_{F_j}}, \quad V_{F_j}=\frac{p_{F_j}}{E_{F_j}}=\frac{\omega_{0j}}{q}$, and $L(s_{j})$ is related to the Lindhard function $\Phi$ by
$$
L_j=L(s_j)=2 \Phi(s_j)=2-s_j \log\left( \frac{s_j+1}{s_j-1}\right), \quad j=p,\,n,\,e.
$$
From the equation of motion of the scalar field given in Eqs. \ref{eomf}, we can write, 
\beq
\left[ \omega^{2}_s - \omega^2 \right] \delta \phi (\vec{q},\omega)
= g_s \sum_{j=p,n} \frac{2 {M^\star}^{(0)}_{j}}{(2\pi)^3}\int \frac{d^3 p}{E^{(0)}_j}\delta f^{(j)},
\eeq
which can be rearranged as following
\beq
\delta \phi (\vec{q},\omega)
= \frac{g_{s} M_{j}^{\star(0)}}{ \omega^2_s - \omega^2  }\sum_{j=p,n}\frac{\delta\rho_j}{E_{Fj}}.
\eeq
Using the above relation in Eq.\ref{eq37}, we obtain
\bwt
\begin{eqnarray}
\delta\rho_{j}&=&\frac{1}{2\pi^2}p_{F j}E^{(0)}_{Fj}L(s_j)\left(\frac{\omega}{q}\delta\mathcal{V}^{(j)}_{z}-\delta\mathcal{V}^{(j)}_{0} +\frac{(g_{s} M_{j}^{\star(0)})^2 }{\omega^{2}_s - \omega^2 }\sum_{i=p,n}\frac{\delta \rho_i}{E^{(0)}_{Fj}E^{(0)}_{Fi}}\right)
\label{master1}
\end{eqnarray}
\ewt
The latter expression will be manipulated in the following to obtain the dispersion relations. First, we remind that from Eq.~(\ref{eq2}) we obtain :
\begin{displaymath}
\delta \mathcal{V}_{\mu}^{(j)}=\left\{\begin{array}{l}
g_{v} \delta V_{\mu}+\tau_{j} \frac{g_{\rho}}{2} \delta \vec{b}_{\mu}+e \frac{1+\tau_{j}}{2} \delta A_{\mu}, j=(p, n)\\
-e \delta A_{\mu}, j=e
\end{array}\right.
\end{displaymath}
Therefore, Eq.~(\ref{master1}), can be written as:
\bwt
\bea
& \delta \rho_{j}=\frac{1}{2\pi^2}p_{F j}E^{(0)}_{Fj}L(s_j)\left[g_{v}\left(\frac{\omega}{q} \delta V_{z}-\delta V_{0}\right)+\tau_{j} \frac{g_{\rho}}{2}\left(\frac{\omega}{q} \delta b_{z}-\delta b_{0}\right)\right.
\cr
&\left. +e \frac{1+\tau_{j}}{2}\left(\left(\frac{\omega}{q}\right)^{2}-1\right) \delta A_{0}
+\frac{\left(g_{s} M_{j}^{\star(0)}\right)^{2}}{\omega_{s}^{2}-\omega^{2}} \frac{1}{E_{F j}^{(0)}} \sum_{i=p, n} \frac{\delta \rho_{i}}{E_{F i}^{(0)}}\right]
\eea
\ewt
Using Eqs.(\ref{fieldt}, \ref{fields}) we can obtain :
\bea
& g_{v}\left(\frac{\omega}{q} \delta V_{z}-\delta V_{0}\right)=\sum_{i=p, n}\left[\left(\frac{\omega}{q}\right)^{2} \frac{g_{v}^{2}}{-\omega^{2}+\omega_{\omega}^{2}-\frac{\xi}{3} V_{0}^{(0)^{2}}}\right. \cr
&-\frac{g_{v}^{2}\left(-\omega^{2}+\omega_{\rho}^{2}\right)}{\left(-\omega^{2}+\omega_{\omega}^{2}\right)\left(-\omega^{2}+\omega_{\rho}^{2}\right)-\left(4 \Lambda_{v} V_{0}^{(0)} b_{0}^{(0)}\right)^{2}} \cr
&+ \left.\frac{4 \Lambda_{v} V_{0}^{(0)} b_{0}^{(0)} g_{v} \frac{g_{\rho}}{2} \tau_{i}}{\left(-\omega^{2}+\omega_{\omega}^{2}\right)\left(-\omega^{2}+\omega_{\rho}^{2}\right)-\left(4 \Lambda_{v} V_{0}^{(0)} b_{0}^{(0)}\right)^{2}}\right] \delta \rho_{i}, \\
& \tau_{j} \frac{g_{\rho}}{2}\left(\frac{\omega}{q} \delta b_{z}-\delta b_{0}\right) =\sum_{i=p, n}\left[\left(\frac{\omega}{q}\right)^{2} \frac{\left(g_{\rho} / 2\right)^{2} \tau_{i} \tau_{j}}{-\omega^{2}+\omega_{\omega}^{2}} \right. \cr
& -\frac{\left(g_{\rho} / 2\right)^{2} \tau_{i} \tau_{j}\left(-\omega^{2}+\omega_{\omega}^{2}\right)}{\left(-\omega^{2}+\omega_{\omega}^{2}\right)\left(-\omega^{2}+\omega_{\rho}^{2}\right)-\left(4 \Lambda_{v} V_{0}^{(0)} b_{0}^{(0)}\right)^{2}} \cr
&+ \left.\frac{4 \Lambda_{v} V_{0}^{(0)} b_{0}^{(0)} g_{v} \frac{g_{\rho}}{2} \tau_{j}}{\left(-\omega^{2}+\omega_{\omega}^{2}\right)\left(-\omega^{2}+\omega_{\rho}^{2}\right)-\left(4 \Lambda_{v} V_{0}^{(0)} b_{0}^{(0)}\right)^{2}}\right] \delta \rho_{i} 
\eea
We use Eq.(\ref{eomf}) to write $\delta A_{0}$ as:
$$
\delta A_{0}=\frac{1}{-\omega^{2}+\vec{q}^{\;2}} \sum_{i=p, e} Q_{i} \delta\rho_{i} \equiv \frac{e}{-\omega^{2}+\vec{q}^{\;2}}\left(\delta\rho_{p}-\delta\rho_{e}\right).
$$
Finally, putting everything together, Eq.(\ref{master1}) becomes:
\bwt
\bea
 \delta \rho_{j}&=&\frac{1}{2\pi^2}\left\{\sum_{i=p, n}\left[\left(\frac{\omega}{q}\right)^{2} \frac{g_{v}^{2}}{-\omega^{2}+\omega_{\omega}^{2}-\frac{\xi}{3} V_{0}^{(0)^{2}}}-\frac{g_{v}^{2}\left(-\omega^{2}+\omega_{\rho}^{2}\right)}{\left(-\omega^{2}+\omega_{\omega}^{2}\right)\left(-\omega^{2}+\omega_{\rho}^{2}\right)-\left(4 \Lambda_{v} V_{0}^{(0)} b_{0}^{(0)}\right)^{2}}\right. \right. \cr
&& +\left(\frac{\omega}{q}\right)^{2} \frac{\left(g_{\rho} / 2\right)^{2} \tau_{i} \tau_{j}}{-\omega^{2}+\omega_{\rho}^{2}}-\frac{\left(g_{\rho} / 2\right)^{2} \tau_{i} \tau_{j}\left(-\omega^{2}+\omega_{\omega}^{2}\right)}{\left(-\omega^{2}+\omega_{\omega}^{2}\right)\left(-\omega^{2}+\omega_{\rho}^{2}\right)-\left(4 \Lambda_{v} V_{0}^{(0)} b_{0}^{(0)}\right)^{2}} \cr
&& +\frac{4 \Lambda_{v} V_{0}^{(0)} b_{0}^{(0)} g_{v} \frac{g_{\rho}}{2} \left(\tau_{i}+\tau_{j}\right)}{\left(-\omega^{2}+\omega_{\omega}^{2}\right)\left(-\omega^{2}+\omega_{\rho}^{2}\right)-\left(4 \Lambda_{v} V_{0}^{(0)} b_{0}^{(0)}\right)^{2}}  \left.+\frac{\left(g_{s} M_{j}^{\star(0)}\right)^{2}}{\omega_{s}^{2}-\omega^{2}} \frac{1}{E_{F i}^{(0)} E_{F j}^{(0)}}\right] \delta \rho_{i} \cr
&& \left.+e \frac{1+\tau_{j}}{2}\left(\left(\frac{\omega}{q}\right)^{2}-1\right) \frac{e}{-\omega^{2}+\vec{q}^{2}}\left(\delta \rho_{p}-\delta \rho_{e}\right)\right\}p_{F j} E^{(0)}_{Fj}L\left(s_{j}\right).
\eea
\ewt
\bea
\delta \rho_{i} &=& - \sum_{j=n, p} F^{i,j} L(s_i) \delta\rho_{j} - L(s_i) C^{i, e}_{A}\delta_{ip} \delta\rho_{e}, \, i=n,\,p \cr
\delta\rho_{e} &=& C^{e, e}_{A} L(s_e) \delta\rho_{e} - C^{e, p}_{A} L(s_e) \delta\rho_{p} 
\label{eqq45}
\end{eqnarray}
where,
$$
F^{i,j}=C^{i,j}_{s} - C^{i,j}_{\omega}-C^{i,j}_{\omega\rho}(\tau_i+\tau_j)-C^{i,j}_{\rho}\tau_i\tau_j -C^{i,j}_{A}\delta_{ip}\delta_{jp},
$$
with, 
\bea
C^{i,j}_{s} &=&  \frac{1}{2\pi^2}\frac{(g_{s} M_{i}^{\star(0)})^2}{\omega^{2} - \omega^2_s }\frac{p_{F_j}}{E^{(0)}_{F_i}} \cr
C^{i,j}_{\omega} &=& \frac{g^2_{v}}{2\pi^2}\left[\frac{\omega^2-\omega^2_{\rho}}{(\omega^2-\omega^2_{\omega})(\omega^2-\omega^2_{\rho}) - (4\Lambda_v V^{(0)}_0 b^{(0)}_0)^2} \right.\cr
&-&\left.\left(\frac{\omega}{q} \right)^2 \frac{1}{\omega^2-\omega^2_{\omega}+\frac{\xi}{3} {V^{(0)}_0}^2 }\right]p_{F_j} E^{(0)}_{F_j} \cr
C^{i,j}_{\rho} &=& \frac{\left(\frac{g_{\rho}}{2}\right)^2}{2\pi^2}\left[
 \frac{\omega^2-\omega^2_{\omega}}{(\omega^2-\omega^2_{\omega})(\omega^2-\omega^2_{\rho})-(4\Lambda_v V^{(0)}_0 b^{(0)}_0)^2} \right.\cr
 &-&\left.\left(\frac{\omega}{q} \right)^2 \frac{1}{\omega^2-\omega^2_{\rho}}\right]p_{F_j} E^{(0)}_{F_j}\cr
C^{i,j}_{\omega\rho} &=& \frac{1}{2\pi^2}\frac{2 \Lambda_v g_{\omega} g_{\rho} V^{(0)}_0  b^{(0)}_0 p_{F_j} E^{(0)}_{F_j}}{(\omega^2-\omega^2_{\omega})(\omega^2-\omega^2_{\rho})-(4\Lambda_v V^{(0)}_0 b^{(0)}_0)^2}\cr 
C^{i,j}_{A} &=&-\frac{1}{2\pi^2}\frac{e^2}{{\vec q}^{~2}}p_{F_j} E^{(0)}_{F_j}.
\eea
Equations \ref{eqq45} are written in terms of the density fluctuations $\delta\rho_{i}$, and they read 
\begin{eqnarray}
\begin{pmatrix}
     1+ F^{pp} L_p    & F^{pn} L_{p} & C^{pe}_{A} L_{p}\\
     F^{np} L_{n}    & 1+ F^{nn} L_n & 0\\
   C^{ep}_{A} L_{e}      & 0 & 1- C^{ee}_{A} L_{e} \\
\end{pmatrix} 
    \begin{pmatrix}
    \delta \rho_{p} \\
    \delta \rho_{n} \\
    \delta\rho_{e}
\end{pmatrix} =0.   
\nonumber
\end{eqnarray}
From the last equation we get the following dispersion relation
\bea
&& \left[1-C^{ee}_A L_e\right]\left[1+L_pF^{pp}+L_n F^{nn}\right. \cr 
&& \left.+ L_p L_n \left(F^{pp}F^{nn}-F^{pn}F^{np}\right)\right] \cr
&& -C^{ep}_A C^{pe}_A L_e L_p\left(1+L_n F^{nn}\right)=0;
\label{detnpe}
\eea
and the density fluctuations are given by
\bea
\frac{\delta\rho_p}{\delta\rho_n}&=& -\frac{F^{pn}L_p }{1+F^{pp}L_p-\frac{C^{p,e}_A L_p C^{e,p}_A L_e}{(1-C^{e,e}_A L_e)}} \cr
\frac{\delta\rho_e}{\delta\rho_p}&=& -\frac{C^{e,p}_A L_e }{1-C^{e,e}_A L_e}.
\eea
In the case of np matter, we remove the electron sector, we obtain  the following dispersion relation:
\beq
1+L_pF^{pp}+L_n F^{nn}+L_p L_n \left(F^{pp}F^{nn}-F^{pn}F^{np}\right)=0,
\label{det1}
\eeq
and in this case the density fluctuation is given by
\beq
\frac{\delta\rho_p}{\delta\rho_n}= -\frac{F^{pn}L_p}{1+F^{pp}L_p}.
\eeq

\begin{figure*}[t]
\includegraphics[width=0.8\linewidth,angle=0]{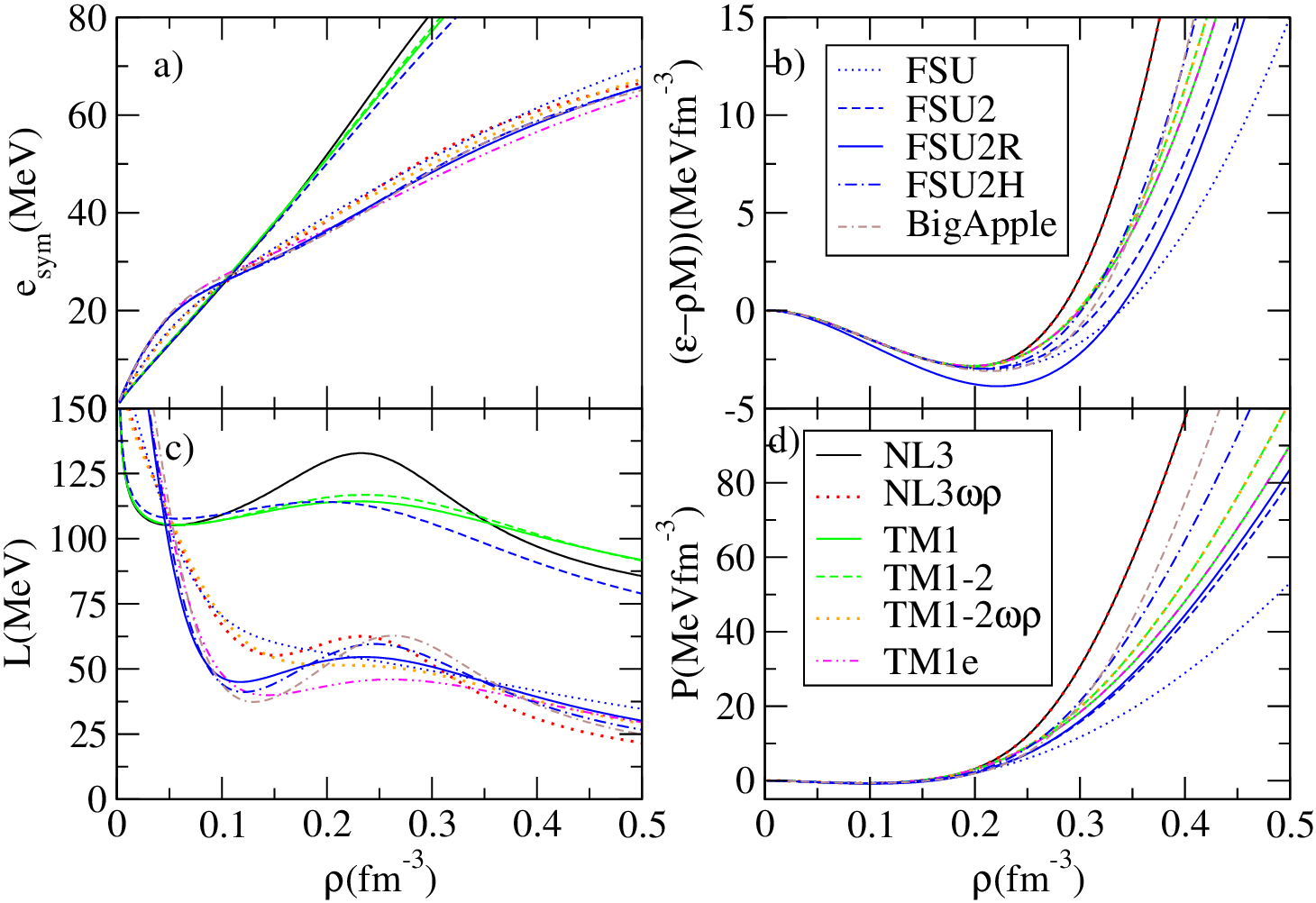}
\caption{Symmetry energy and symmetry energy slope as function of baryon density (left panel), energy density and pressure as a function of baryon density (right panel), for symmetric nuclear matter.}
\label{ener_sym}
\end{figure*}

\section{Results and discussion}

In this section we present and discuss the collective longitudinal modes of nuclear matter, which is composed of protons and neutrons and which we refer to as np matter and $\beta$-equilibrium neutron star matter, which forms the outer core of neutron stars and is composed of neutrons, protons, and electrons and which we refer to as npe matter. For simplicity, neither muons nor hyperons will be considered in this discussion. However, muons generally appear below the saturation density, while hyperons could probably appear at the baryonic densities considered although it is not certain. We will first introduce the RMF models that will be analysed.

\begin{table*}[htb]
 \begin{tabular}{lcccccccc}
\hline\hline
  & $\rho_0$ & $B$ & $K$ & $Q_0$ &$E_\mathit{sym}$   &$L$ & $K_{0,\hbox{sym}}$ & $Q_{0,sym}$ \\
& $ [\mathrm{fm}^{-3}]$& [MeV]& [MeV]& [MeV]&[MeV]& [MeV] &  [MeV] &  [MeV] \\\hline 
NL3 &0.148&-16.24&271 & 197.9&37.4&118 &100.5 & 182.4  \\
NL3$\omega\rho$&0.148&-16.24&271& 197.9&31.5&55& -8.0 & 1397.0 \\
TM1-2&0.145&-16.3&281& -200.9 & 36.9& 111 & 42.0 & -31.6 \\
TM1-2 $\omega\rho$&0.146&-16.3&281 & -197.0 &32.1& 55 & -70.4 & 1069.5 \\
BigApple & 0.155 & -16.344 & 227 & -203.9 & 31.3 & 39.8 & 88.8 & 1122.5  \\
FSU & 0.148 & -16.30 & 230 & -521.5 & 32.6 & 60.5 & -51.4 & 426.6  \\
FSU2&0.1505&-16.28&238 & -149.2 &37.6&113 & 25.4 & -165.9\\
FSU2R&0.1505&-16.28&238& -135.5&30.7 & 47 &55.86 & 190.3\\
FSU2H&0.1505&-16.28&238&-246.7 &30.5 &44.5 & 86.9 & 652.4\\
TM1 &0.145&-16.26&281& -284.9&36.8& 110 & 33.5 & -65.4\\
TM1e & 0.145 & -16.3& 281 & -284.9 & 31.4& 40 &3.1 & 848.8  \\
\hline
\end{tabular}
\caption{Nuclear matter properties of the RMF models considered in this study \label{tab:nuclear}
}
\end{table*}
In the following, the results are given for two sets of models:
set I (NL3, NL3$\omega\rho$, TM1-2, TM1-2$\omega\rho$, and BigApple) and set II (FSU, FSU2, FSU2R, FSU2H, TM1, and TM1e). Models belonging to set I either do not have the $\omega^4$ term in the Lagrangian density (NL3 and NL3$\omega\rho$) or the coupling in front of this term is small, and as a result are much stiffer at high densities.  In Fig.~\ref{ener_sym},
the symmetry energy and its slope  (left panels)  and the energy density and the pressure (right panels)   are represented as function of the nuclear density.

Models NL3, TM1, TM1-2 and FSU2 are characterized by a large slope of the symmetry energy at saturation  density, above 100 MeV.  Relative to the other models, they have a smaller symmetry energy at sub-saturation densities and a larger one above the saturation density. This behavior affects the reaction of the system to small isovector like perturbations as will be discussed below.
The other models  FSU, FSU2R, FSU2H, TM1e, TM1-2$\omega\rho$, BigApple and NL3$\omega\rho$ have a quite large symmetry energy at sub-saturation densities, but show a soft behavior at supra-saturation densities.  Except to FSU2R, all models show a similar energy density and pressure for symmetric nuclear matter below saturation density. Above this density, the chosen models span a quite large region in the pressure (energy density) baryonic density space. Models that show the stiffest behavior at high densities, NL3, NL3$\omega\rho$, BigApple and FSU2H are precisely the ones that have the smallest values of the $\xi$ parameter, see table \ref{tab:parameters}.

\begin{figure}[h]
\includegraphics[width=0.95\linewidth,angle=0]{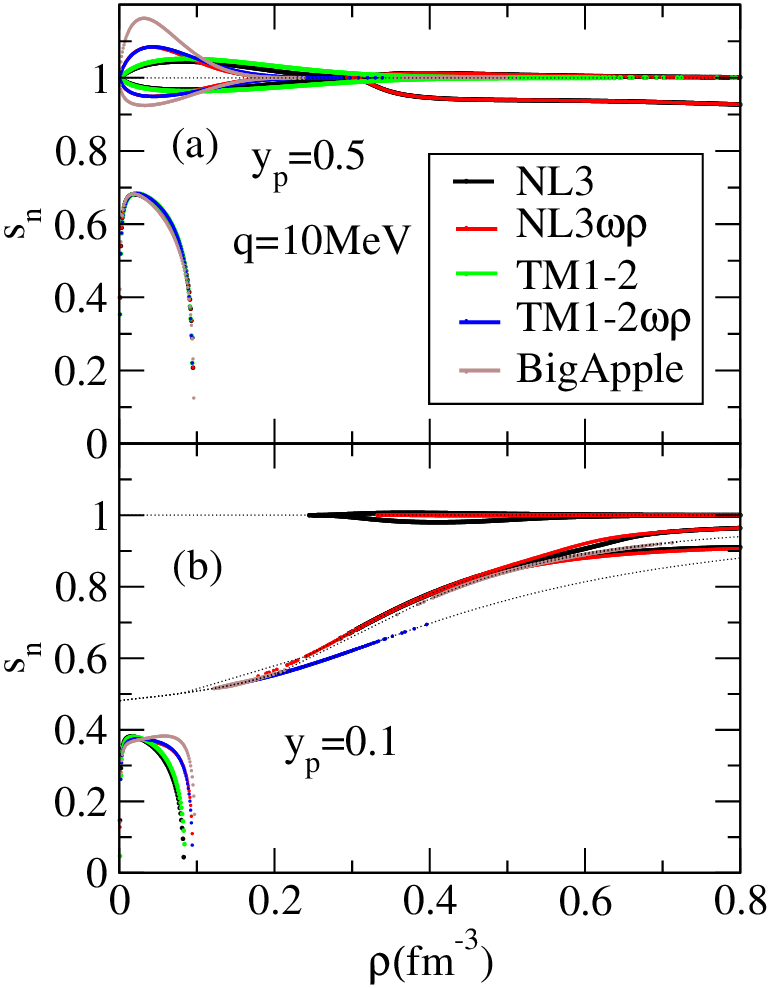}
\caption{(Color online) Nuclear collective modes $s_{n}=\omega/(q \, V_{F_n})$ for $q=10$MeV and $y_p=0.5$ (upper panel) $y_p=0.1$ (lower panel), and for NL3, NL3$\omega \rho$, TM1-2, TM1-2$\omega \rho$, BigApple models. The Fermi velocity of neutrons (dotted thin line $s_n=1$) and protons (dotted thin lines).}
\label{figure1a}
\end{figure}

\begin{figure}[h]
\includegraphics[width=0.995\linewidth,angle=0]{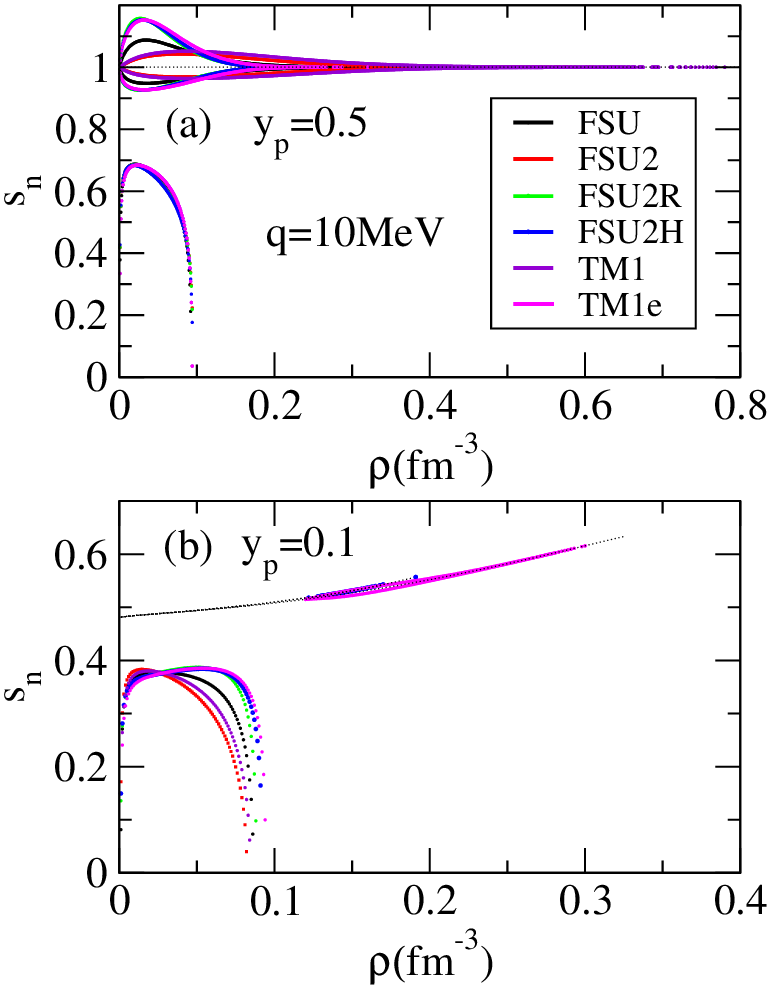}
\caption{(Color online) Nuclear collective modes $s_{n}=\omega/(q \, V_{F_n})$ for $q=10$~MeV and $y_p=0.5$ (upper panel) and  $y_p=0.1$ (lower panel), for FSU, FSU2, FSU2R, FSU2H, TM1 and TM1e models. The Fermi velocity of neutrons (dotted thin line $s_n=1$) and protons (dotted thin lines) is also plotted.}
\label{figure1b}
\end{figure}

In the following we discuss the results obtained from Eq.~(\ref{detnpe}) and Eq.~(\ref{det1}). Nuclear collective modes are obtained from the solution of the dispersion relation given in Eq.~(\ref{det1}).
As discussed in Ref.~\cite{avancini05}, the isoscalar and isovector collective modes are identified by the ratio of the proton to neutron density fluctuations for a given nuclear density, $\delta \rho_p/\delta\rho_n$. Isoscalar modes are characterized by a positive ratio, $\delta \rho_p/\delta\rho_n>0$, while isovector modes correspond to $\delta \rho_p/\delta\rho_n<0$. With this definition, protons and neutrons move in phase for isoscalar modes and out of phase for isovector modes.

In Fig.~\ref{figure1a} we display the longitudinal isoscalar and isovector modes divided by $\omega_{0i}=q\, V_{Fi}$  ($s_i=\omega/(q\, V_{F_i})$) for (a)  symmetric nuclear matter, $y_p=0.5$, and (b) for asymmetric matter with $y_p=0.1$, for the momentum transfer equal to $q=10$~MeV and for models of the set I. We also include the lines corresponding to  $\omega_{0i}$ (thin dotted lines) that identify the Fermi levels and allow us to identify the modes that do not suffer Landau damping, i.e. collective  zero sound modes above the Fermi level that do not couple to single particle-hole excitations.  

\begin{figure}[ht]
\includegraphics[width=1\linewidth,angle=0]{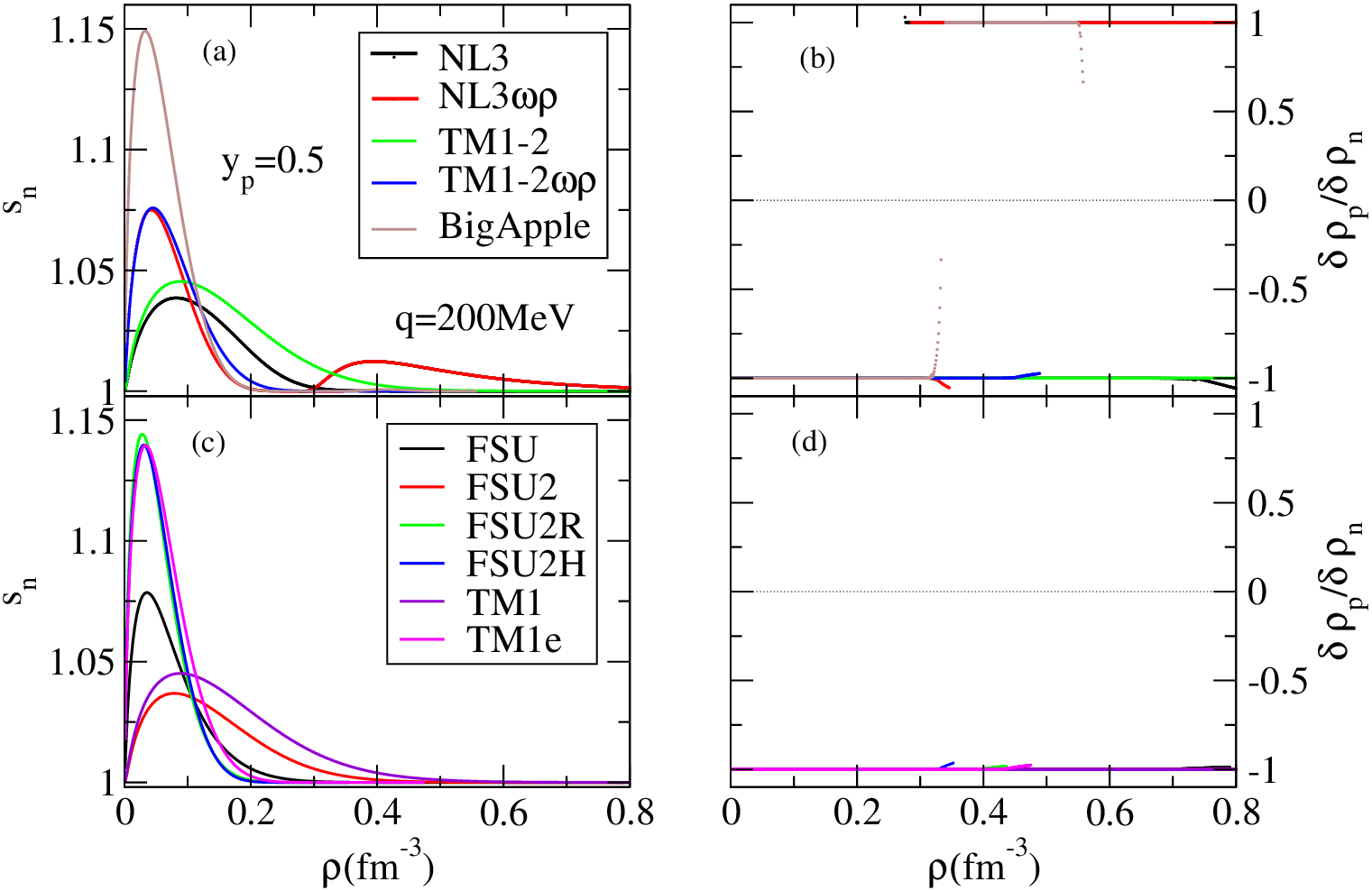}
\caption{(Color online) Collective modes as a function of the baryonic density for $y_p=0.5$, for q =200 MeV, for set I model (top panel) and set II models (bottom).}
\label{figure2as}
\end{figure}

In Fig.~\ref{figure1a}(a), three different sets of modes for all models considered are identified: i) a mode that emerges alone below the Fermi velocity limit and for densities below $\rho\sim 0.1$ fm$^{-3}$. This is an isoscalar like mode that defines the instability region of symmetric nuclear matter; ii) a set of pairs of modes below a density $\rho\sim 0.2-0.4$fm$^{-3}$ depending on the model. The lower limit was obtained for the models  NL3$\omega\rho$, TM1-2$\omega\rho$ and BigApple  with a smaller symmetry energy slope. NL3 and TM1-2 modes extend to 0.3 and 0.5 fm$^{-3}$, respectively. Below saturation density, BigApple model attains the largest magnitudes, followed by the NL3$\omega\rho$ and TM1-2$\omega\rho$ models, which achieve almost half the magnitude of the BigApple model, while the NL3 and  TM1-2 models show  the smallest magnitudes. These models are characterized by a different slope of the symmetry energy  $L$ at saturation,  BigApple having the smallest and NL3 and TM1-2 the largest. This implies that below  (above) saturation density the symmetry energy is largest (smallest) for BigApple and smallest (largest) for NL3 and TM1-2. The magnitude is larger for a larger symmetry energy. The modes extend to larger densities in the models with larger symmetry energy above saturation density. The properties of these modes are determined by the density behavior of the symmetry energy and are isovector like modes, the protons and neutrons move out of phase; iii) a set of pairs of modes above a density  0.2-0.3 fm$^{-3}$ again depending on the models. These modes are isoscalar like, with protons and neutrons moving in phase and they exist only for the models with a zero or very small $\xi$ coupling, NL3, NL3$\omega\rho$ and BigApple. The properties of these modes are determined by the density dependence of the symmetric nuclear matter energy and only models with a stiff EOS allow for these modes. For symmetric nuclear matter these modes coincide for the models  NL3 and NL3$\omega\rho$ because these two sets of models are coincident for symmetric nuclear matter and only differ for asymmetric matter. 
For the BigApple model the mode lies between 0.35~fm$^{-3}$ to 0.5fm$^{-3}$, because the EOS is very soft around saturation density, it has the smallest $K$, but becomes stiffer than all the other models with a non zero $\omega^4$ term above 0.35~fm$^{-3}$.  The isoscalar model requires a stiff EOS to exist. Notice that the isoscalar mode disappears at large densities, i.e. for a large enough proton and neutron Fermi energy.

In Fig.~\ref{figure1a}(b) the modes present in asymmetric matter with a proton fraction 0.1 are plotted together with  the limit defined by the Fermi velocity of neutrons (black thin dotted line at $s_n=1$) and protons (black thin dotted line). For NL3 and NL3$\omega\rho$, there are still four modes, with one above and one below each of these lines. However, for TM1-2$\omega\rho$ and BigApple, there are only  two modes, one above and the other one below the proton Fermi velocity lines. This is because the mode disappears at large Fermi momenta and  in very asymmetric matter the neutron Fermi momentum  of neutrons is too large. 
At low densities, the isoscalar mode that defines the instability region is also present. No isovector mode occurs for this proton fraction.

In Fig.~\ref{figure1b}, we show the same as Fig.~\ref{figure1a} using the set II models. For symmetric nuclear matter, see  Fig.~\ref{figure1b}(a), no isoscalar like modes appear at high densities. Only  the pair  of isovector modes exist at lower densities, together   with an isoscalar mode at low densities which comes alone and   defines the instability region.  The models
FSU2R, FSU2H and TM1e produce isovector modes with the highest amplitudes in the density range between 0 and 0.25 fm$^{-3}$; FSU2 and TM1 have isovector modes with the smallest amplitudes which lie in the density interval between 0 and 0.5 fm$^{-3}$ and the  FSU model has an intermediate behavior. This can be understood as before, the models with the stiffest symmetry energy below or above the saturation density have the largest modes, in the respective range of a stiffer symmetry energy. The extension of the mode is also defined by the stiffness of the symmetry energy at high densities, with greater stiffness corresponding to a larger density range.

From Fig.~\ref{figure1b}(b) and comparing with the Set I models, we see that the neutron and proton modes disappear for most of the models and only for the TM1e and FSU2H models the proton modes persist. The isoscalar mode at low densities, which defines the instability region, is still present. 

\begin{figure}[t]
\includegraphics[width=1\linewidth,angle=0]{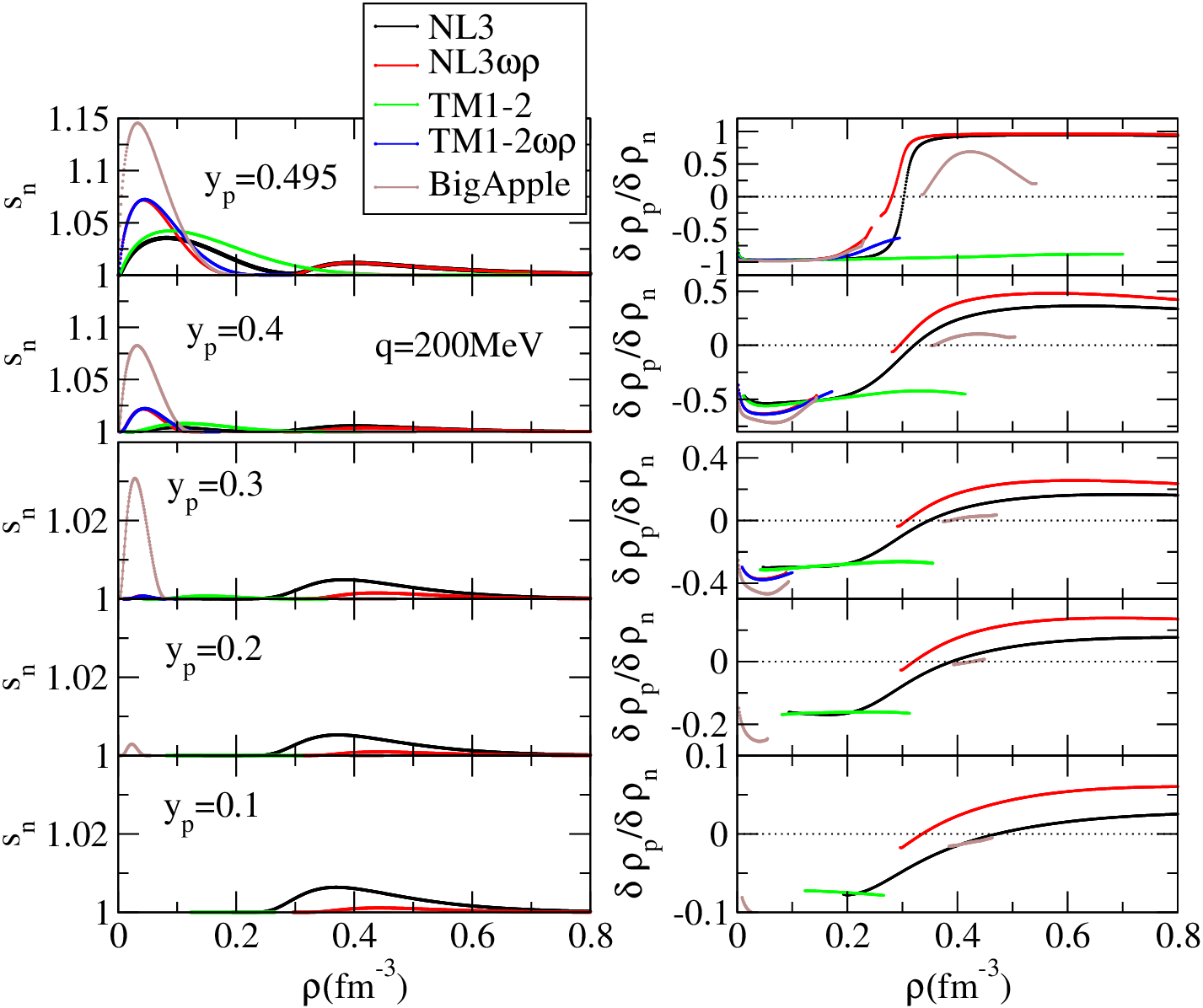}
\caption{(Color online) Collective modes (left panels) and ratio between the proton and neutron amplitudes  $\delta \rho_p/\delta\rho_n$ obtained for $q=200$~MeV in terms of the baryonic density for $y_p=0.495, 0.4, 0.3, 0.2, 0.1$, and for NL3, NL3$\omega \rho$, TM1-2, TM1-2$\omega \rho$, BigApple models.}
\label{figure2a}
\end{figure}

In order to better understand the results shown in Figs.~\ref{figure1a} and ~\ref{figure1b}, in the following we discuss the stable (undamped) collective modes with a speed of sound above the neutron Fermi speed for different proton fractions, and, in particular, their dependence on the isospin asymmetry. 

In Fig.~\ref{figure2as}, these modes are represented (left panels) together with the corresponding ratios $\delta \rho_p/\delta\rho_n$ (right panels) for the set I models (top panel) and set II models (bottom panel), with q = 200 MeV and for symmetric nuclear matter, $y_p=0.5$.  These are the curves represented above the Fermi level in   Fig.~\ref{figure1a}(a) and Fig.~\ref{figure1b}(a), top panels. The right panels identify them as pure isoscalar or pure isovector modes. The pure isoscalar modes are present only above approximately two times saturation density and only exist in models with a stiff EOS at high densities, as discussed before. Let us now understand the effect of introducing isospin asymmetry.

\begin{figure}[t]
\includegraphics[width=1\linewidth,angle=0]{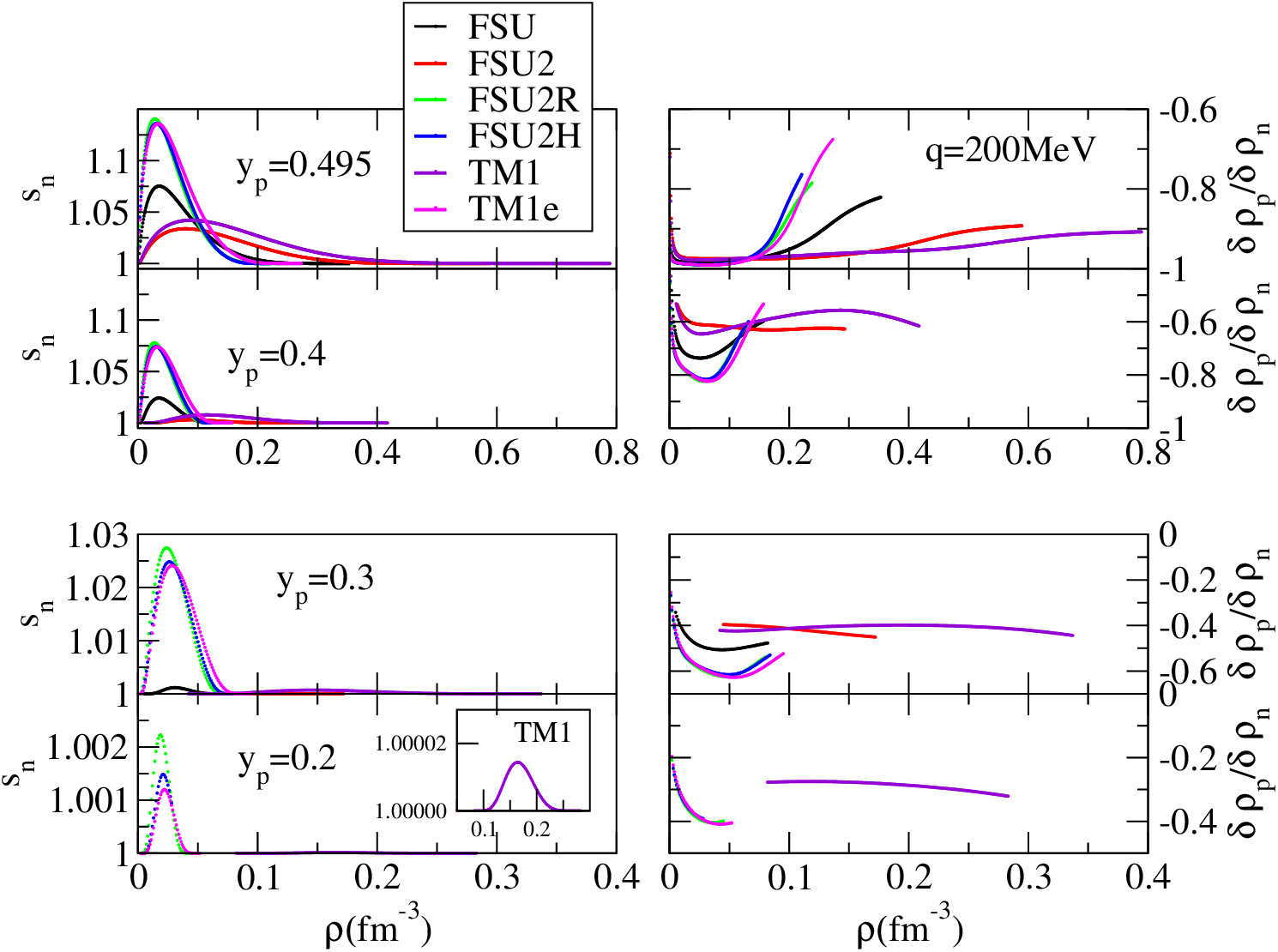}
\caption{(Color online) Collective modes (left panels) and ratio between the proton and neutron amplitudes  $\delta \rho_p/\delta\rho_n$ obtained for $q=200$~MeV in terms of the baryonic density for $y_p=0.495, 0.4, 0.3, 0.2$, and for FSU, FSU2, FSU2R, FSU2H, TM1 and TM1e models.}
\label{figure2b}
\end{figure}

In Fig.~\ref{figure2a}, for the set I models,  $q = $200 MeV and decreasing proton fractions, we plot the sound velocity of the collective modes (left panels)  and the corresponding ratio of proton to neutron transition densities (right panels). The same for the set II models is given in Fig.~\ref{figure2b}. For symmetric nuclear matter the isoscalar and isovector modes are completely decoupled as discussed in Ref.~\cite{avancini05}. However, for $y_p<0.5$ these modes are coupled. For NL3, the mode is represented by a continuous line that changes character from isovector to isoscalar at a given baryonic density designated by  transition density. The transition to a isoscalar like mode occurs only  for NL3, NL3$\omega\rho$ and BigApple models, due to the stiff EOS at high densities. 
Previous studies \cite{greco2003, avancini05} have discussed the identification of a transition density, but they did not consider the possibility that this transition may not occur, depending on the model's properties. The transition density increases with decreasing proton fraction, and this behavior is model dependent.
Note, however, that while NL3 has a continuous mode from the low densities isovector like mode to the high densities isoscalar like mode, the other models have separated branches, and at intermediate densities the mode may not exist.
 For TM1-2 and TM1-2$\omega\rho$ the  mode has always isovector character. The same occurs for the set II models, with a softer behavior at large densities; these models only have a isovector like  mode which disappears for a proton fraction $y_p < 0.2$.  A proton fraction above 0.2 may occur during the neutrino trapped phase of a protoneutron star.
 
\begin{figure}[t]
\includegraphics[width=0.95\linewidth,angle=0]{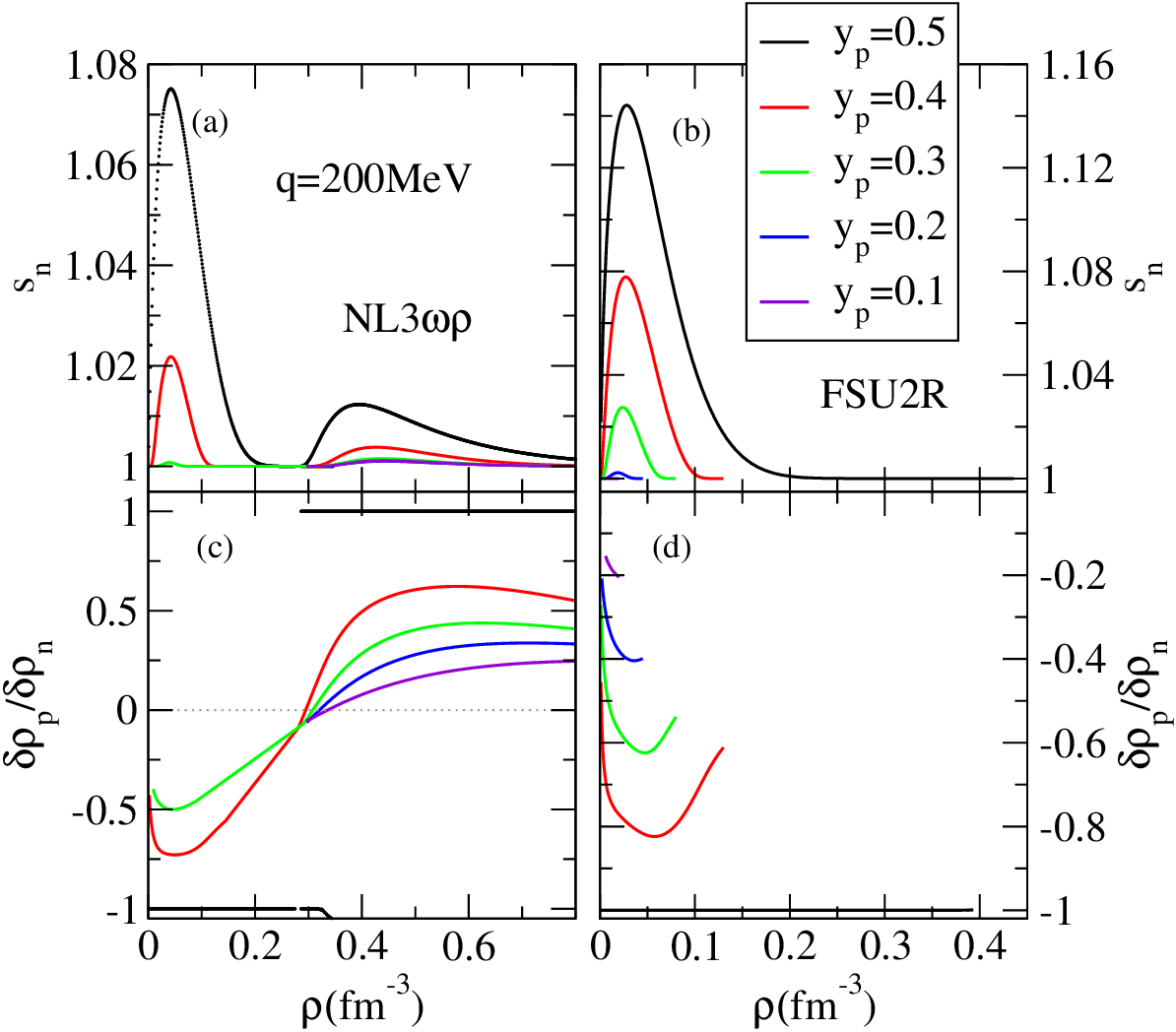}
\caption{(Color online) Collective modes (top panel) and corresponding proton to neutron transition density ratio (bottom panel) as function of the baryonic density, for different proton fractions $y_p$, with $q=200$~MeV and for NL3$\omega \rho$ model (left panels) and FSU2R model (right panels).}
\label{fig11}
\end{figure}

The effect of the isospin asymmetry on the collective mode character, is more clearly seen in Fig.~\ref{fig11} where the collective modes (top panel) and corresponding proton to neutron transition density ratio (bottom panel) for a momentum transfer of  $q=200$~MeV are shown   as function of the baryonic density, for different proton fractions $y_p$, and for NL3$\omega \rho$ model (left panels) and FSU2R model (right panels). For the NL3$\omega \rho$ model,   the collective mode changes its character from isovector to isoscalar at the transition density $\sim 0.3$ fm$^{-3}$ for $y_p\geqslant 0.3$. For $y_p < 0.3$ the mode only exists for densities above the transition density when it is isoscalar like. 
For FSU2R the mode exists only as an isovector like mode at densities below $\sim 0.2$ fm$^{-3}$, extending to smaller and smaller densities as the proton fraction decreases, and it disappears for $y_p < 0.1$, see Fig. \ref{figure2b}. 
If the asymmetry is too large the repulsive contribution of the isovector channel is too large, and there is not enough attraction for the mode to be excited.
Note, however, that below $\rho\lesssim\rho_0/2$ lies the inner crust of the NS and  matter is not homogeneous. The present results should only be considered for the core matter, the inner crust requesting a different approach. 

\begin{figure*}[t]
\includegraphics[width=0.9\linewidth,angle=0]{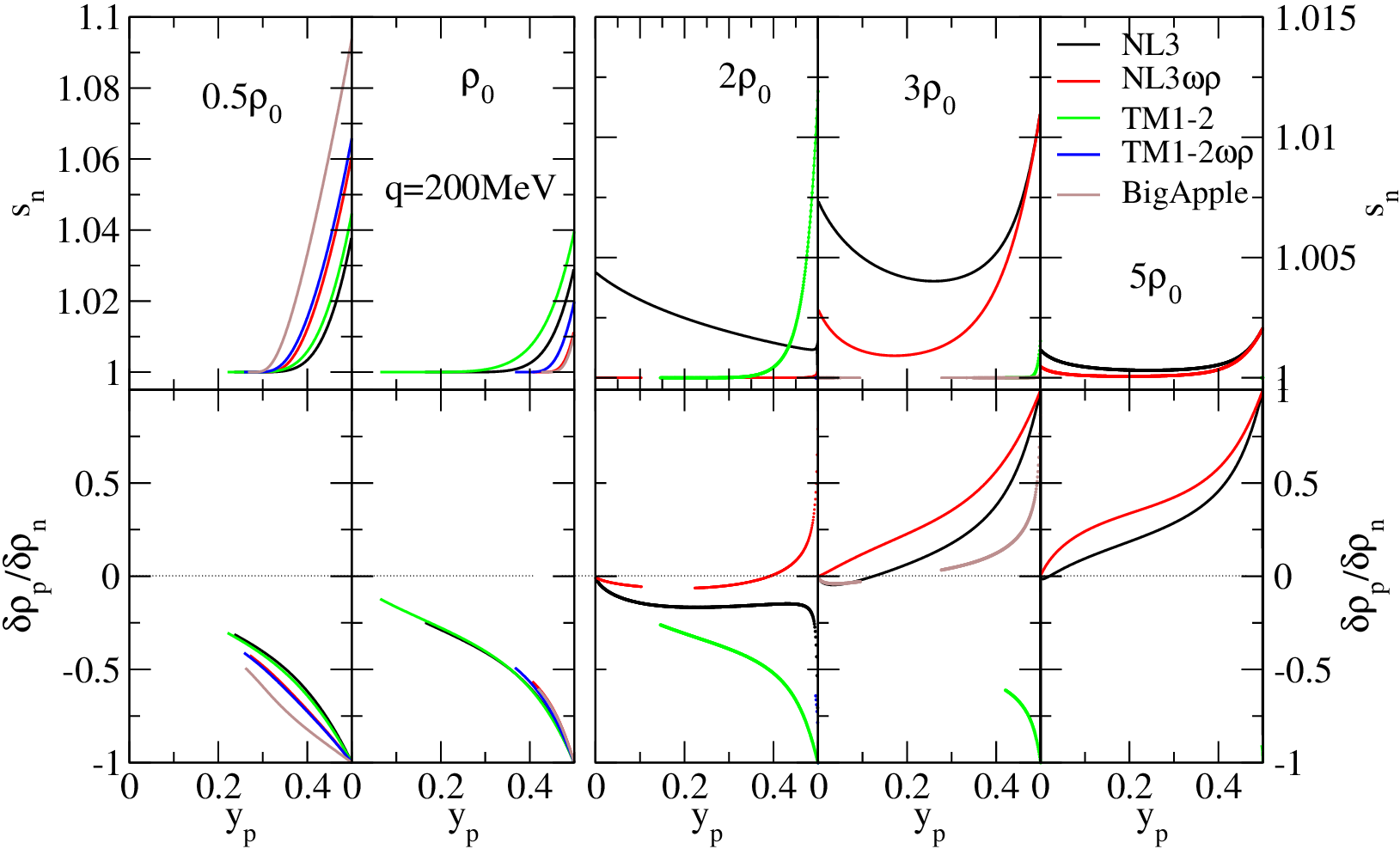}\\
\includegraphics[width=0.9\linewidth,angle=0]{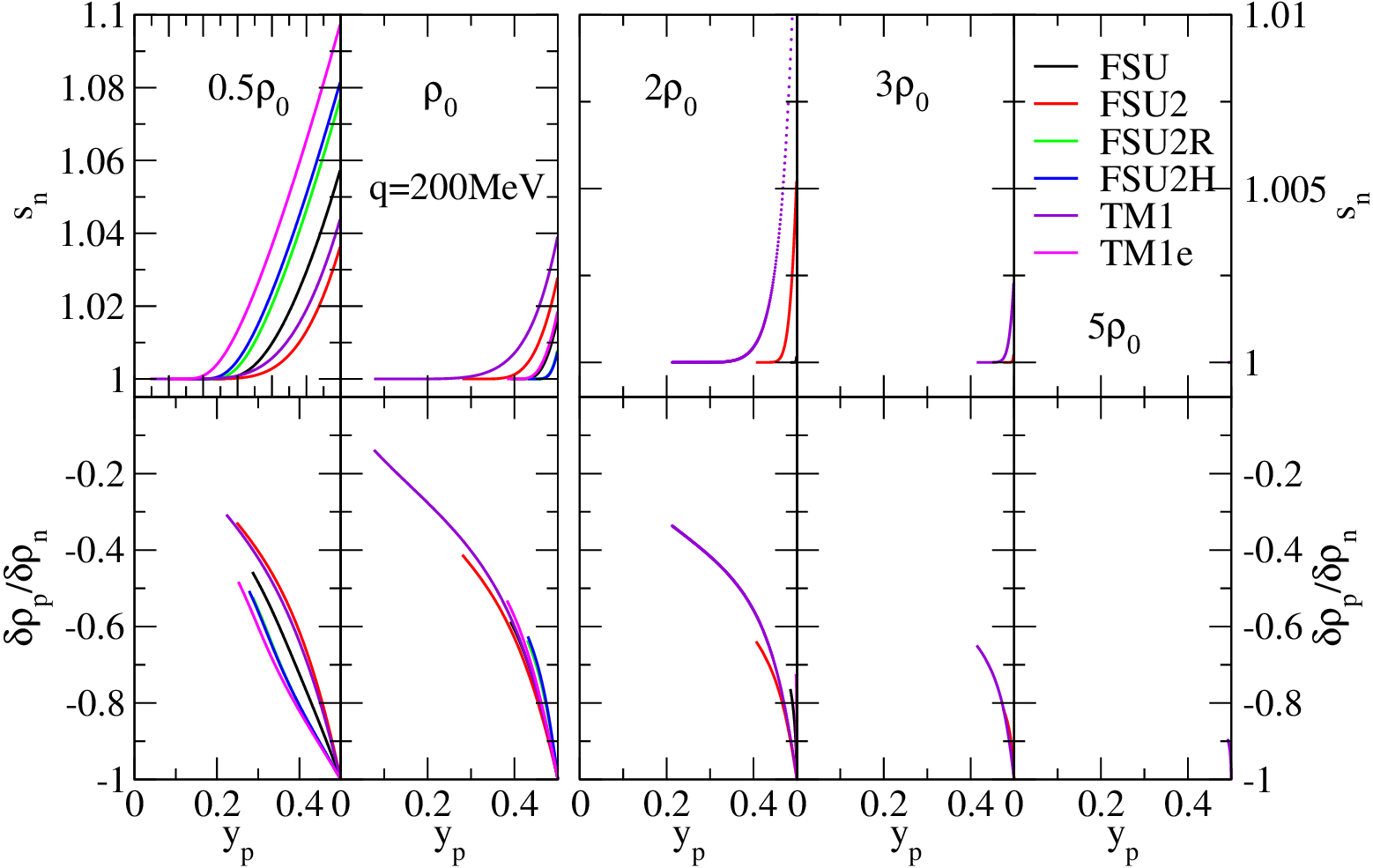}
\caption{(Color online) Collective modes in function of the isospin asymmetry for different densities and corresponding proton to neutron transition density ratio, for $q=200$MeV and the models  NL3, NL3$\omega \rho$, TM1-2, TM1-2$\omega \rho$ models (top), and   FSU, FSU2, FSU2R, FSU2H, TM1 and TM1e models (bottom).}
\label{figure4_05sb1}
\end{figure*}

In Fig.~\ref{figure4_05sb1} 
we have plotted for all the models the collective modes and corresponding ratio of proton to neutron density fluctuations  as a function of the proton fraction at different baryonic densities, in order to compare the behavior of the different models.

The representative  densities $\left\{0.5\rho_0, \rho_0, 2\rho_0, 3\rho_0, 5 \rho_0\right\}$ were considered together with the momentum transfer $q = 200$ MeV, for the set I  models 
(Fig.~\ref{figure4_05sb1} top) and set II models (Fig.~\ref{figure4_05sb1} bottom). These figures summarize Figs. \ref{figure2a} and \ref{figure2b}. As discussed before, some models of set I  present both modes with an isovector character at low densities and isoscalar character at high densities.  The isovector mode at low densities disappears for proton fractions below 0.28 to 0.07, depending on the model ($y_p\leqslant 0.24$ for NL3; at $y_p\leqslant 0.27$ for NL3$\omega\rho$, TM1-2$\omega\rho$ and BigApple models; and at $y_p\leqslant 0.22$  for TM1-2).  The differences are directly related with the strength of the isovector channel: a weak coupling does not originate the necessary restoring force if the asymmetry is large. Below $\rho=0.1$fm$^{-3}$, BigApple has the largest symmetry energy within set I, and, among the two sets is one of the models with the largest subsaturation symmetry energy and coupling $g_\rho$. 
The extension of the density range where the mode exists also depends on the magnitude of the symmetry energy. The non-linear $\omega^2 \rho^2$ term  whose magnitude is defined by the coupling $\Lambda_\omega$ weakens the isovector channel and at 2$\rho_0$ or above only the models with $\Lambda_\omega$ equal to zero or very small have a non-zero isovector mode: these are essentially the models NL3, TM1, TM1-2 and FSU2. Some of the other models may still have this mode if the proton fraction is close to 0.5.  At twice and three times the saturation density, the isovector mode of TM1-2, TM1 and FSU2 still survive with similar characteristics, but NL3 and NL3$\omega\rho$ (and BigApple at 3$\rho_0$) show a different behavior: the mode involves essentially only the neutrons with just a few protons independently of the proton fraction, and may change to isoscalar like mode above a given density. This is the result of a stiff isoscalar behavior, and this explains why the BigApple also shows this mode although in limited conditions. In set II, bottom panels, no model shows these characteristics: all models have a soft EOS at large densities, with a $\xi$ coupling $\gtrsim 0.01$. 
\begin{table*}[htb]
 \begin{tabular}{l|cccccccccc}
 \hline
\hline
Model   & $\rho$ & y$_p$ &$p_{F_n}$& $p_{F_p}$   &$\omega_{0e}$& M$^{*}$/M & V$_{Fn}$ & V$_{Fp}$ & N0$_{n}$  & N0$_{p}$ \\
   & & (\%) & (MeV)&(MeV)&(MeV)  & &  & & (1/GeVfm$^{3}$) & (1/GeVfm$^{3}$) \\
\hline
 Set I model \\
\hline    
NL3  & $\rho_0$   &  8.07 & 313.98 & 139.52 & 7.76     &    0.603    &    0.485    &    0.239   &     2.34    &    1.04    \\
& 2$\rho_0$  & 19.31 & 378.76 & 235.14 & 13.09    &    0.309    &    0.793    &    0.629   &     1.45    &    0.90   \\
& 3$\rho_0$    & 24.18 & 424.66 & 290.15 & 16.15     &    0.189    &    0.922    &    0.853   &     1.00    &    0.68  \\
\hline     
NL3$\omega\rho$  & $\rho_0$    &  5.87 & 316.46 & 125.49 &  6.984      &    0.604    &    0.487    &    0.216    &    2.37    &    0.94 \\
& 2$\rho_0$    & 10.89 & 391.50 & 194.28 & 10.81       &    0.319    &    0.794    &    0.544    &    1.55    &    0.77    \\ 
& 3$\rho_0$    & 13.35 & 443.98 & 238.05 & 13.25      &    0.199    &    0.922    &    0.787    &    1.09    &    0.59   \\ 
\hline        
TM1-2 & $\rho_0$    &  7.90  & 312.03 & 137.61 &  7.66    &    0.640    &    0.461    &    0.223    &    2.47    &    1.09    \\
& 2$\rho_0$         &  18.16 & 377.96 & 228.81 & 12.73    &    0.407    &    0.704    &    0.514    &    1.90    &    1.15    \\
& 3$\rho_0$         &  23.54 & 422.96 & 285.59 & 15.89    &    0.282    &    0.848    &    0.734    &    1.48    &    1.00    \\
\hline        
TM1-2$\omega\rho$ & $\rho_0$    & 6.01 & 314.15 & 125.64 &  6.99       &    0.638    &    0.466   &     0.206    &    2.49    &    1.00     \\       
& 2$\rho_0$    &  10.13 & 389.94 & 188.35 & 10.48      &    0.410    &    0.713   &     0.441    &    1.98    &    0.96    \\     
& 3$\rho_0$    &    12.85 & 441.82 & 233.42 & 12.99     &    0.289    &    0.853   &     0.654    &    1.58    &    0.84   \\    
\hline          
BigApple & $\rho_0$    &    5.82 & 321.44 & 127.05 &  7.07       &    0.617    &    0.485   &     0.214    &    2.46    &    0.97    \\     
& 2$\rho_0$    &  10.09 & 398.76 & 192.36 & 10.71       &    0.329    &    0.791   &     0.529    &    1.62    &    0.78   \\      
& 3$\rho_0$    &    12.96 & 451.55 & 239.35 & 13.32     &    0.198    &    0.925   &     0.790   &     1.11    &    0.59   \\    
\hline
Set II model \\          
\hline  
FSU & $\rho_0$    &   5.28 & 317.12 & 121.18 &  6.74     &    0.616    &    0.480    &    0.205   &     2.42   &     0.92   \\
   & 2$\rho_0$    &   8.95 & 394.32 & 181.97 & 10.13     &    0.444    &    0.687    &    0.400   &     2.17   &     1.00   \\
\hline         
FSU2 & $\rho_0$    &   8.10 & 315.70 & 140.50 &  7.82    &    0.600    &    0.489     &   0.242   &     2.34    &    1.04    \\
    & 2$\rho_0$    &  17.71 & 383.38 & 229.76 & 12.79   &    0.375    &    0.736     &   0.546   &     1.78    &    1.07   \\
\hline        
FSU2R &  $\rho_0$    &    5.54 & 318.61 & 123.76 &  6.89    &     0.601   &     0.492    &    0.214   &     2.37    &    0.92  \\
     &  2$\rho_0$    &    9.72 & 395.41 & 188.08 & 10.47   &     0.379   &     0.744    &    0.468   &     1.85    &    0.88   \\    
\hline         
FSU2H &  $\rho_0$    &   5.50 & 318.65 & 123.48 &  6.87   &     0.601    &    0.491    &    0.214    &    2.37    &    0.92  \\
      & 2$\rho_0$    &  10.01 & 394.99 & 189.94 & 10.57  &     0.338    &    0.779    &    0.513    &    1.65    &    0.80   \\          
\hline       
TM1 & $\rho_0$    &    7.90 & 312.03 & 137.61 &  7.66     &     0.640    &    0.461    &    0.223    &    2.47    &    1.09    \\
    & 2$\rho_0$   &   17.99 & 378.23 & 228.09 & 12.69   &     0.419    &    0.693    &    0.502    &    1.96    &    1.18   \\
\hline 
TM1e & $\rho_0$    &     6.02 & 314.15 & 125.67 &  6.99   &     0.641    &    0.463    &    0.205    &    2.49    &    1.00   \\
    & 2$\rho_0$    &     9.11 & 391.41 & 181.79 & 10.12   &     0.425    &    0.700    &    0.415    &    2.06    &    0.96 \\
\hline
\end{tabular}
\caption{Model, proton fraction $y_p$, Fermi momenta $p_{F_{i}}$, $q=0$ electron plasmon frequencies $\omega_{0e}$ in $\beta$-equilibrium neutron star matter, effective mass, Fermi velocities $V_{F_{i}}$, and level densities for three different baryonic densities. The values given for the proton fraction $y_p$ are obtained by microscopic calculation for set I and set II models; these values are adopted for the Figs. \ref{fig_vsqnl3}, \ref{fig9_set1} and \ref{fig9_set2}. 
\label{tableII}}
\end{table*}

\begin{figure}[t]
\includegraphics[width=0.95\linewidth,angle=0]{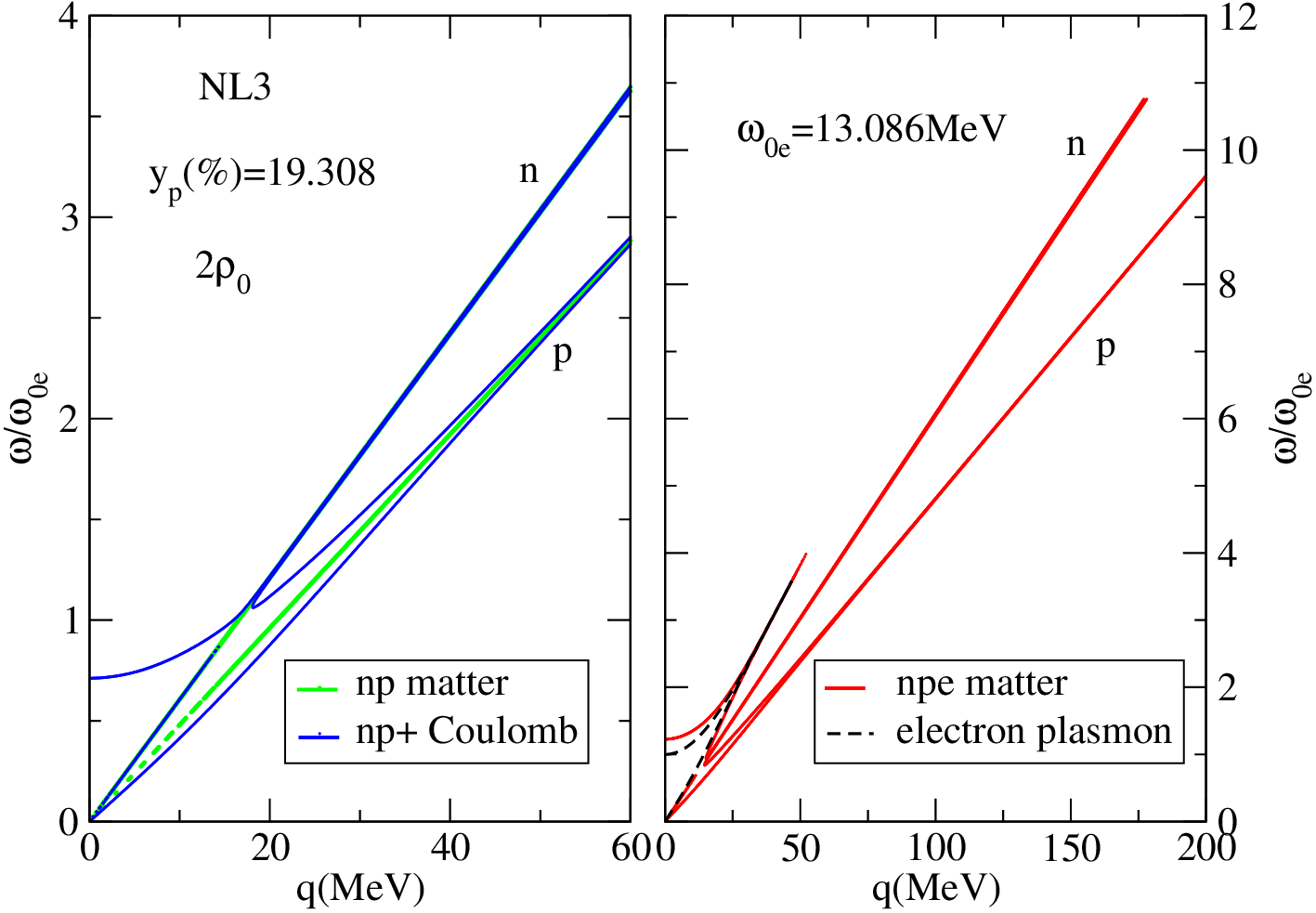}\\
\includegraphics[width=0.95\linewidth,angle=0]{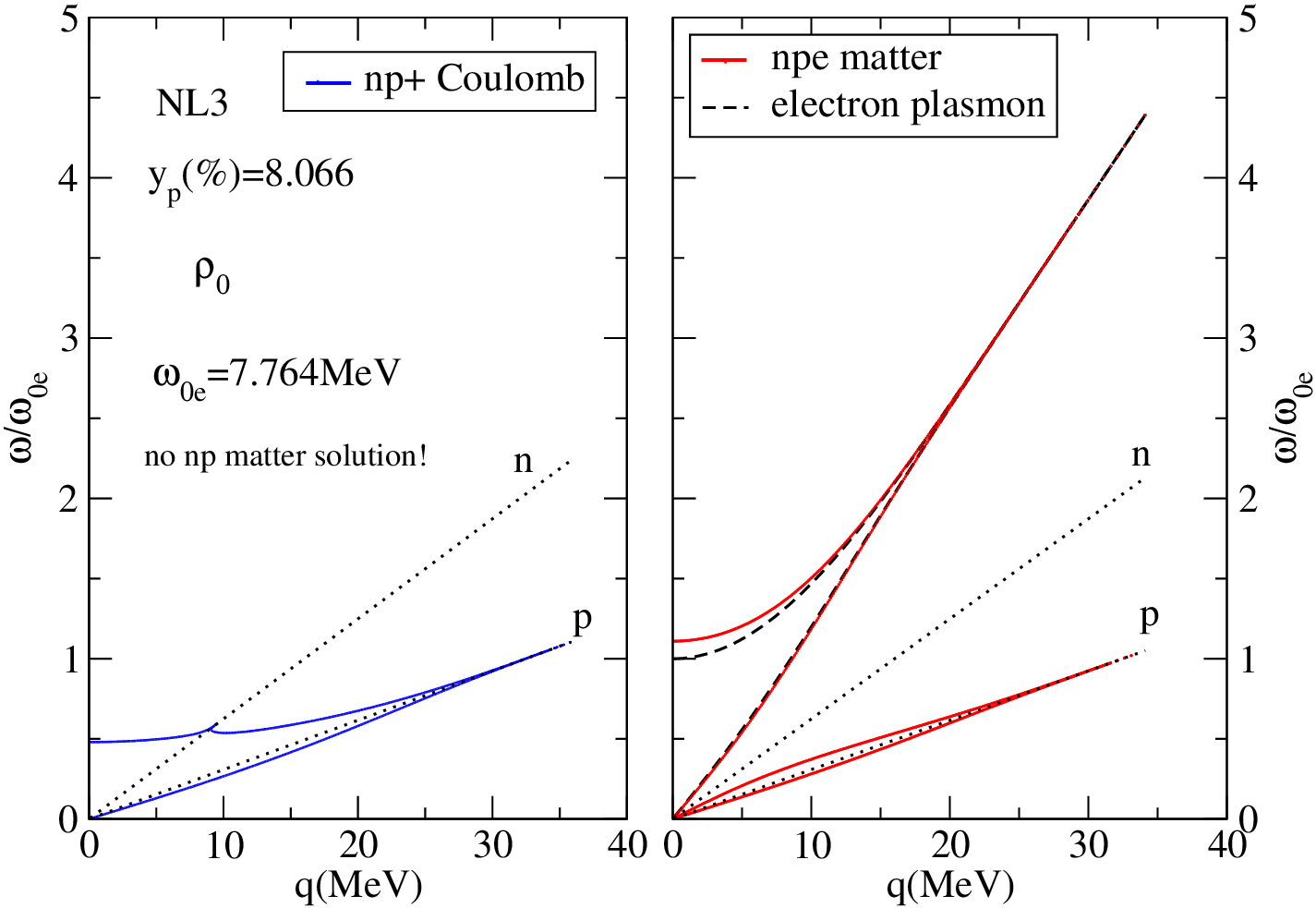}
\caption{Collective modes as a function of the momentum transfer $q$ and for the $\beta$-equilibrium proton fraction  (top) $y_p(\%)=19.31$ at 2$\rho_0$ and (bottom)  $y_p(\%)=8.07$ at $\rho_0$ using the NL3 model. Green solid lines are for np matter neglecting the Coulomb interaction, blue solid lines are for np including the Coulomb effect, red solid lines for npe matter and black dashed lines are for electrons in a positive background.  The thin dotted lines represent $qV_{F_i}$ for $i=n,\, p$.  Only the nuclear like modes propagate for momenta equal to 200 MeV or above. For $y_p(\%)=8.07$ at $\rho_0$ no np modes propagate neglecting the Coulomb field. }
\label{fig_vsqnl3}
\end{figure}

\begin{figure*}[htb]
\includegraphics[width=0.9\linewidth,angle=0]{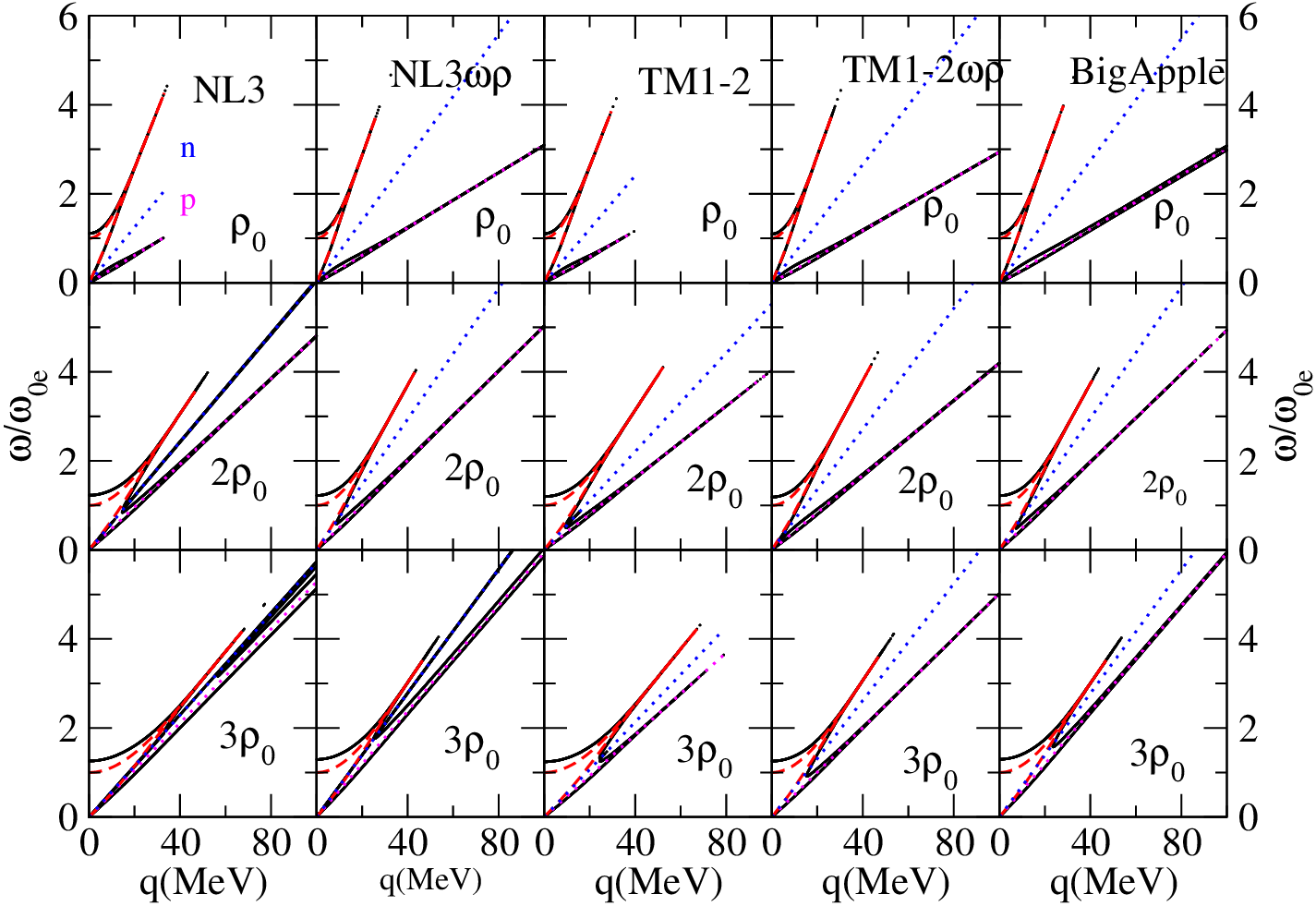}
\caption{
Collective modes as a function of the momentum transfer $q$ for $\beta$-equilibrium neutral matter described by set I models.  Matter at  three different densities, respectively, $\rho_0$, $2\rho_0$ and $3\rho_0$ is considered. The corresponding proton fractions, the  proton and neutron Fermi momenta and the  $q=0$ electron plasmon frequency  are   given in  Table \ref{tableII}. Results for npe matter are represented by black lines, for a relativistic gas of free electrons by red dashed lines. The Fermi velocity of neutrons (blue dotted lines) and protons (magenta dotted lines).}
\label{fig9_set1}
\end{figure*}

\begin{figure*}[htb]
\includegraphics[width=0.9\linewidth,angle=0]{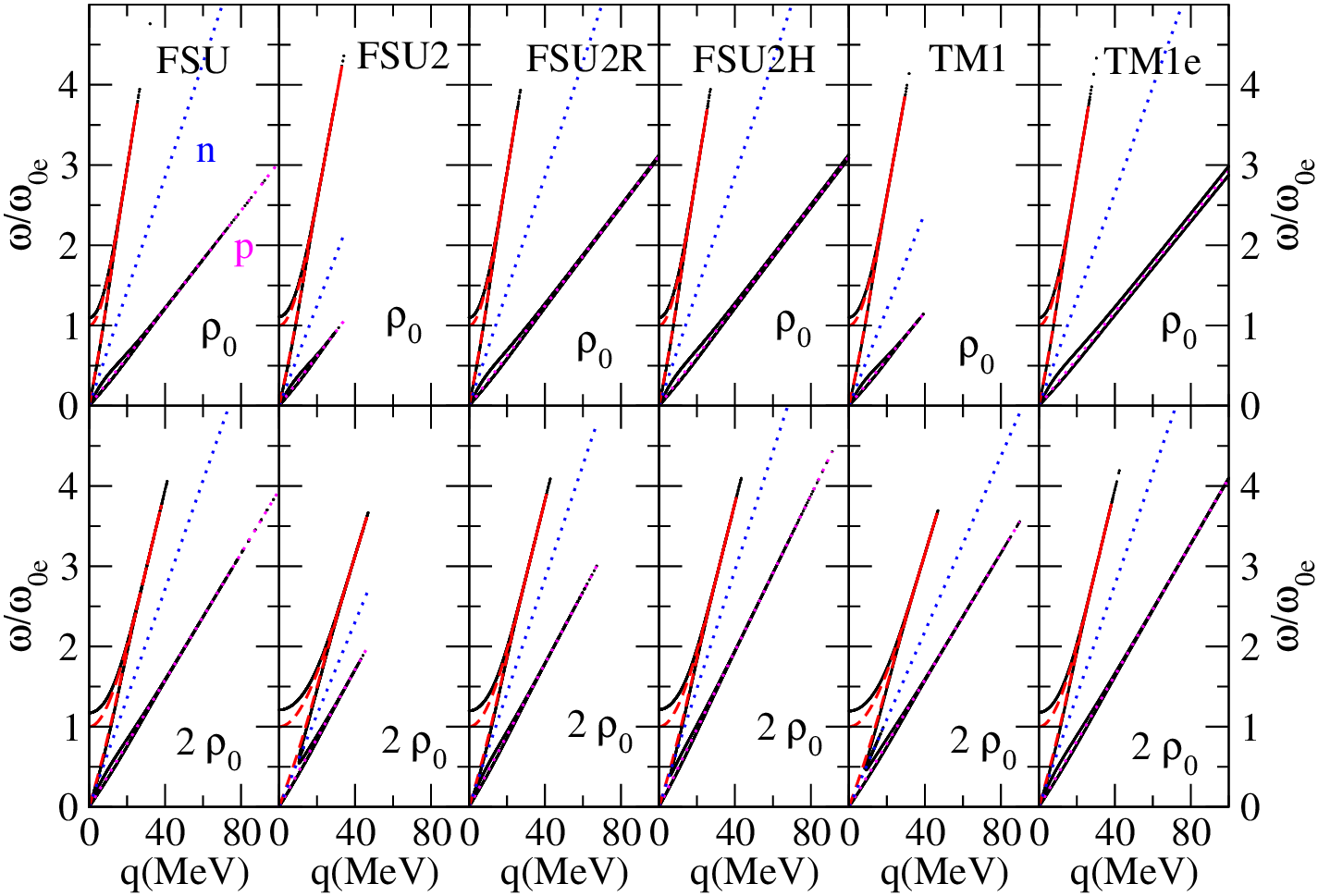}
\caption{Collective modes as a function of the momentum transfer $q$ for $\beta$-equilibrium neutral matter described by set II models.  Matter at  two different densities, respectively, $\rho_0$ and $2\rho_0$ are considered. The corresponding proton fractions, the  proton and neutron Fermi momenta and the  $q=0$ electron plasmon frequency  are   given in  Table \ref{tableII}. Results for npe matter are represented by black lines, and  for a relativistic gas of free electrons by red dashed lines. The Fermi velocity of neutrons (blue dotted lines) and protons (magenta dotted lines). }
\label{fig9_set2}
\end{figure*}

In the following, in order to complete our investigation, we consider normal neutron star matter, excluding the case of superfluid matter. 
Taking into account the range of densities where the nuclear modes are none zero, we  compare the behavior of the frequency of the collective modes as a function of the momentum transfer at $\rho=\rho_0, \, 2\rho_0,\, 3\rho_0 $ for models of set I, and  at $\rho=\rho_0,\, 2\rho_0 $ 
for models of set II. In theses ranges of density neutron star matter is expected to be homogeneous and not affected by possible 
strange components like hyperons. The proton fraction is determined by the $\beta$ equilibrium condition and varies according to the equation of state (EOS) of each model. Table~\ref{tableII} provides the proton fraction as a function of baryon density for the selected values of baryonic density, along with the corresponding Fermi momenta and electron plasma frequency (at $\rho_e = \rho_p$), effective mass, Fermi velocity, and the density of state of each nucleon. Although the results are presented for zero temperature, the analysis can be easily extended to finite temperatures. 

If we ignore the nucleon degrees of freedom, the dispersion relation Eq.~\ref{detnpe} simplifies to
\beq
1-C^{ee}_A L(s_e)=0.
\label{plasmon}
\eeq
We recall that Jancovici \cite{jancovici1962} has studied the longitudinal response of a relativistic degenerate electron gas in 1962. The left-hand side of the last equation determines the dielectric constant of the electron gas under the conditions $p<<p_{Fe}$ and $\omega << E_{Fe}$, when quantum recoil terms are negligible and the Vlasov equation can be applied. To analyze how nuclear modes couple with plasmon modes as given by  Eq.~(\ref{plasmon}), we will use dashed lines in the figures that follow to represent the response of the free electron gas when applicable.  For the electrons in a positive background (the jelly model), there are two modes to be considered: a sound-like mode and a plasmon mode with a frequency.
$$
\omega_{0e}=\sqrt{\frac{e^2\rho_e}{E_{Fe}}},
$$
at transfer momentum $q=0$MeV.

Before discussing the behavior of the different modes, we summarize in  Fig.~\ref{fig_vsqnl3} for the NL3 model  in  $\beta$-equilibrium matter at 2$\rho_0$ with a proton fraction $y_p(\%)=19.31$ (top panels), and at $\rho_0$ with a proton fraction $y_p(\%)=8.07$ (bottom panels), the possible collective modes as a function of the transferred momentum for different matter scenarios. The left panels refer to np matter with no Coulomb interaction (solid green lines) and np matter including the Coulomb field and considering a negative uniform  background (solid blue lines), while the right panels refer to
electrons in a positive uniform  background (thin  black dashed lines) and neutral npe matter in $\beta$-equilibrium  (solid blue lines). Thin dotted lines identify the $qV_{F_i}$ lines for neutrons and protons.  We first discuss the modes that propagate at 2$\rho_0$: i) for electron matter we identify the plasmon and a sound like mode; ii) for np matter neglecting the proton charge two pairs of sound like modes above and below the lines $qV_{F_i}, \, i=n,p$ are obtained; iii)  if the Coulomb field is introduced one of the proton modes is a plasmon like mode and the other a strongly damped sound mode; iv) considering npe matter, a plasmon like mode above the electron plasmon is identified, together with two pairs of sound like modes, that lie above and below the lines $qV_{F_i}$. Some of these modes do not propagate  at low momenta. As discussed in \cite{Baldo:2008pb}, for npe matter the proton plasmon  mode suffers the effect of electron screening and behaves as a sound like mode.  Note that at low momentum the electrons and protons are strongly coupled and the npe modes deviate from the np modes. As momentum increases this coupling becomes weaker and the modes tend to the electron plasmon mode (top mode)  or to the np modes (middle and bottom modes). 
At $\rho_0$ no modes propagate in np matter neglecting the Coulomb interaction, and when the Coulomb interaction is introduced the protons give rise to a plasmon like mode in the uniform negative background. When npe matter is considered the electron plasmon appears at high energy and the proton plasmon is screened and transformed into a sound like mode. No neutron like mode propagates at this density.
The sound like modes are the modes identified in Fig. \ref{figure1a} and \ref{figure1b} above and below the curves $qV_{F_i}$.
In the following, we will compare the response  of the different models  of the two sets at three densities as a function of momentum.

Figs.~\ref{fig9_set1} and \ref{fig9_set2} depict, for $\beta$-equilibrium neutral npe matter with a high isospin asymmetry, the dependence of the energy of the collective modes on the momentum transfer $q$, respectively, for set I (for $\rho=\rho_0,\; 2\rho_0$, and $3\rho_0$) and set II
(for $\rho=\rho_0,\; 2\rho_0$) models (black lines). We also include the results for a relativistic gas of free electrons with the same electron density as the npe matter (red dashed lines), and the lines $qV_{F_i},\, i=p,\,n$. We distinguish two types of modes: plasmon like modes and sound like nuclear modes. The plasmon mode is essentially an electron mode. As already seen above, for a momentum above $q\sim$10-20 MeV the plasmon mode does not mix  with the nuclear modes at low densities when the proton fraction is small. 
{Moreover, beyond a  momentum $q\sim 30-40$ MeV, the plasmon-like modes do not propagate in either family of models. This is expected because the Coulomb interaction varies with $\sim 1/q^2$ and, therefore, as $q$ increases it becomes weaker. For high enough values of $q$ this contribution becomes negligible 
and  Eq.~\ref{plasmon} has no solution.

All models have a proton-like zero sound mode. This mode generally propagates for $q>100$ MeV or even above 200 MeV, although for models with a stiff symmetry energy like TM1, TM1-2, NL3, FSU2 it only propagates for $q\lesssim 40$ MeV for $\rho=\rho_0$. At this density the proton fraction is small, of the order of ~7\%. At 2 $\rho_0$ a similar behavior is obtained.
For models like FSU2R, FSU2H, and TM1e,  which have a small symmetry energy and, therefore, smaller proton fraction, the mode may propagate well beyond $q=200$ MeV. FSU has an intermediate behavior. 

At twice saturation density, the NL3 model, with approximately a fraction of protons equal to 20\%,  exhibits two nuclear modes (neutron-like and proton-like), with the proton mode still propagating at $q=200$ MeV, and the neutron mode  propagating  up to $q=180$ MeV. The same occurs at three times saturation density for NL3 and NL3$\omega\rho$.
These are the models with the stiffest EOS at high densities. The neutron-like mode is the mode  present in the left panels of Fig. \ref{figure2a} at $\sim 0.3$fm$^{-3}$. 

TM1-2$\omega\rho$ and NL3$\omega\rho$ (at $\rho_0$)  models do not exhibit neutron-like modes but  the proton-like mode  propagates up to $q=200$ MeV due to their soft symmetry energy. On the contrary, the TM1-2 and BigApple models have  a much stiffer symmetry energy and, therefore, the  proton-like modes do not propagate beyond $q\sim 100-120$ MeV.

\section{Conclusions and outlooks}

In the present work, we have applied the covariant formulation of the Vlasov equation presented in \cite{Heinz,avancini2018} to determine the collective modes of hadronic matter
within a relativistic mean-field (RMF) approach. By using the covariant Vlasov approach, we were able to obtain expressions for dispersion relations that precisely coincided with those obtained by the generator method in Refs.~\cite{avancini05,cp2006b}. We have discussed the modes that propagate both in nuclear matter with a fixed proton fraction and in $\beta$-equilibrium matter formed by neutrons protons and electrons, within a set of RMF models frequently considered in the literature to describe neutron star matter. These models are characterized by several properties that distinguish them such as the stiffness/softness of the EOS for symmetric nuclear matter and the symmetry energy. It is shown that the propagation of excitations is very sensitive to these properties.

Two different types of modes have been identified: isoscalar like modes and isovector  like modes, corresponding, respectively, to protons and neutrons moving in phase and out of phase. It was shown that the possible propagation of  both types of collective modes depends strongly on the model: stable isoscalar modes occur above $\sim 2\rho_0$ only if the EOS is stiff enough. Models that include the $\omega^4$ term become too soft at high densities and the isoscalar mode does not propagate.  Note that in \cite{greco2003} the authors have always encountered this mode because all the models considered were very stiff.  In our study the only models that are similar are NL3 and NL3$\omega\rho$. These two models are often considered in the literature. NL3 has been fitted to the properties of nuclear matter but it has a quite large incompressibility as well as the symmetry energy at saturation and its slope. The  NL3$\omega\rho$ model has a softer symmetry energy but keeps the NL3 stiffness at high densities. In \cite{Fortin:2016hny} it was shown that NL3 fails several nuclear matter properties, while NL3$\omega\rho$ only fails some constraints imposed by chEFT calculations. However, neither of these models predicts the tidal deformability and or the radius of 1.4$M_\odot$ star within  the values obtained from observations  see \cite{Fortin:2016hny,Malik:2018zcf}. Similar conclusions to the ones drawn in our study concerning the relation of the high density stiffness and the isoscalar zero sound mode has also been drawn in \cite{Ye:2023fhy}. In this study, the authors verify the correlation between the strength of the $\omega^4$ coupling and the disappearance of the zero sound mode.

Below $\sim 2\rho_0$ an  isovector mode appears in all models. Its magnitude and range of densities and momenta for which it  propagates depend on the stiffness of the symmetry energy.  A stiffer symmetry energy favors its appearance, however the mode does not propagate at high momenta. Models that include the $\rho^2\omega^2$ term have large symmetry energy below saturation density and therefore, a strong isovector mode in this range of densities. As soon as the non-linear terms becomes non-negligible the symmetry energy softens and the mode disappears. 

The effect of the Coulomb field was also discussed, both considering np  nuclear matter and npe neutron star matter. In the absence of the Coulomb field, two sets of sound like modes may appear above and below the line defining the proton and neutron Fermi level.
Including the Coulomb field, we find that in  npe matter the most energetic  mode is always the electron plasmon mode, which, however, does not propagate for a momentum above $\sim 40-50$ MeV. The presence of electrons and inclusion of Coulomb field affects the propagation of the nuclear modes at low momenta, and the proton plasmon like mode behaves as a sound like mode as discussed in~\cite{Baldo:2008pb}.

In summary,  through the analysis of several popular nuclear matter models, the present study  provides some  qualitative insights into the behavior of nuclear matter and stellar structure when subjected to perturbations that originate density fluctuations. The models considered describe quite successfully nuclear matter by including nonlinear meson terms. A different family of models is the density-dependent relativistic mean-field (DD-RMF) family, which will also be addressed in the future. We also intend to include temperature effects to assess their impact on both stable and unstable collective modes. This addition will help us understand the role of thermal effects, which could be significant in systems like neutron stars or heavy-ion collisions, where both density and temperature play a critical role in determining the properties of nuclear matter. The collective transverse modes will also be studied using the covariant Vlasov approach.

\section*{ACKNOWLEDGMENTS}
This work was partially supported by funds from FCT (Fundação para a Ciência e a Tecnologia, I.P, Portugal) under projects UIDB/04564/2020 and UIDP/04564/2020, with DOI identifiers 10.54499/UIDB/04564/2020 and 10.54499/UIDP/04564/2020, respectively, and the project 2022.06460.PTDC with the associated DOI identifier 10.54499/2022.06460.PTDC. %
\bibliographystyle{apsrev4-1}
\bibliography{bibliography}

\end{document}